\documentclass[preprint,5p,twocolumn,10pt]{elsarticle}

\usepackage[pdftex,hypertexnames=false,linktocpage=true]{hyperref}
\hypersetup{colorlinks=true,linkcolor=red,anchorcolor=darkgreen,citecolor=green!50!blue,filecolor=darkgreen,urlcolor=green,bookmarksnumbered=true,pdfview=FitB}

\usepackage{graphicx}

\usepackage{array}
\usepackage{booktabs}
\usepackage{arydshln}
\usepackage{multirow}

\def\arraystretch{1.8}

\newcommand{\ie}{\textit{i.e.}} 
\newcommand{\fig}[1]{Fig.~(#1)}
\newcommand{\figs}[1]{Figs.~(#1)}
\newcommand{\tab}[1]{Table~#1}

\newcommand{\eq}[1]{Eq.~(#1)}
\newcommand{\eqs}[1]{Eqs.~(#1)}
\newcommand{\paper}[1]{Ref.~#1}
\newcommand{\papers}[1]{Refs.~#1}

\usepackage{amsmath,amssymb,mathtools} 
\ifpdftex
    \usepackage{bbm}
    \newcommand{\mybb}[1]{\mathbbm{#1}}
    
    \newcommand{\mycal}[1]{\mathcal{#1}}
\else
    \usepackage[bold-style=ISO]{unicode-math}
        \newcommand{\mybb}[1]{\symbb{#1}}
        
        \newcommand{\mycal}[1]{\symcal{#1}}
\fi

\usepackage{siunitx}  
\makeatletter
\newcommand{\vast}[1]{\bBigg@{#1}}
\makeatother

\newcommand{\intd}[3][]{\ifthenelse{\isempty{#3}}{\mathrm{d}^{#1} #2}{\frac{\mathrm{d}^{#1} #2}{#3}}\;}
\newcommand{\rmd}{\mathrm{d}}

\usepackage{stackengine}

\newcommand{\cond}{\; | \;}
\newcommand{\setcond}[2]{\ifthenelse{\isempty{#2}}{\{#1\}}{\{#1\cond{}#2\}}}

\usepackage{braket}
\usepackage{cancel}
    
\newcommand{\gev}[1]{{#1\,\mathrm{GeV}}}
\newcommand{\gevSq}[1]{{#1\,\mathrm{GeV}^2}}

\newcommand{\LCm}{{-}} 
\newcommand{\LCp}{{+}}

\newcommand{\LCperp}{{\perp}}

\newcommand{\nmax}{N_{\mathrm{max}}}

\let\ppSq\perpSq
\let\ppAbs\perpAbs

\newcommand{\fI}[1]{f_1^{#1}}
\newcommand{\fITperp}[1]{f_{1T}^{\perp #1}}

\newcommand{\gIL}[1]{g_{1L}^{#1}}
\newcommand{\gIT}[1]{g_{1T}^{#1}}
\newcommand{\hI}[1]{h_{1}^{#1}}

\newcommand{\hITperp}[1]{h_{1T}^{\perp #1}}
\newcommand{\hILperp}[1]{h_{1L}^{\perp #1}}
\newcommand{\hIperp}[1]{h_1^{\perp #1}}

\let\fITp\fITperp

\usepackage{braket}

\graphicspath{{../../plots/}}

\begin{document}

\begin{frontmatter}
\title{Transverse momentum structure of proton within the basis light-front quantization framework}

\author[ucas,imp,keylab]{Zhi Hu\corref{cor1}}
\ead{huzhi@impcas.ac.cn}

\author[ucas,imp,keylab]{Siqi Xu\corref{cor1}}
\ead{xsq234@impcas.ac.cn}

\author[ucas,imp,keylab]{Chandan Mondal\corref{cor1}}
\ead{mondal@impcas.ac.cn}

\author[ucas,imp,keylab]{Xingbo Zhao\corref{cor1}}
\ead{xbzhao@impcas.ac.cn}

\author[iowa]{James P. Vary\corref{cor1}}
\ead{jvary@iastate.edu}

\author[]{\\\vspace{0.2cm}(BLFQ Collaboration)}

\address[ucas]{School of Nuclear Physics, University of Chinese Academy of Sciences, Beijing, 100049, China}
\address[imp]{Institute of Modern Physics, Chinese Academy of Sciences, Lanzhou, Gansu, 730000, China}
\address[keylab]{CAS Key Laboratory of High Precision Nuclear Spectroscopy, Institute of Modern Physics, Chinese Academy of Sciences, Lanzhou 730000, China}
\address[iowa]{Department of Physics and Astronomy, Iowa State University, Ames, IA 50011, USA}

\cortext[cor1]{Corresponding author}

\begin{abstract}
We obtain the leading-twist valence quark transverse-momentum-dependent parton distribution functions (TMD PDFs) for the proton within the basis light-front quantization (BLFQ) framework. Our results are consistent with lattice QCD calculations and our previous results for the collinear limit. We also obtain consistency with the Soffer-type bounds. Within our approach, we find that six T-even TMDs in the leading twist are all independent of each other, and previously found model-dependent relations do not hold. This is a promising sign that our results are representative of future, more extensive treatments of QCD. Furthermore, we obtain a non-trivial $ x $-dependence of the $ \braket{\ppSq{p}} $ and some consistency with the Gaussian ansatz but only in the small $ \ppSq{p} $ region. Those features suggest our results may be a useful alternative in future experimental extractions.
\end{abstract}
\end{frontmatter}

\section{Introduction}
Recently, there are numerous theoretical investigations and experiments~\cite{ZEUS:2003pwh,COMPASS:2008isr,HERMES:2009lmz} aimed to understand the transverse momentum dependent parton distributions (TMD PDFs, or simply TMDs in the following) and the generalized parton distributions (GPDs), which encode the three-dimensional information of the quark in the nucleon. The TMDs~\cite{Collins1985,Collins2011} are necessary to describe the Semi-Inclusive Deep Inelastic Scattering (SIDIS)~\cite{Bacchetta2007,Ji2005} or Drell-Yan processes~\cite{Tangerman:1994eh,Collins:2002kn,Zhou:2009jm}, while the GPDs~\cite{Diehl2003,Belitsky:2005qn} are required for exclusive processes like deeply virtual Compton scattering (DVCS)~\cite{Ji:1996nm,Goeke:2001tz} or vector meson productions~\cite{Goloskokov2008,Collins1997}.

The TMDs are the extended version of collinear parton distribution functions (PDFs), capturing the three-dimensional structural information of hadrons in momentum space. These distributions also encode the knowledge about the correlations between spins of the target and momenta of the partons. At leading twist, there are eight TMDs for the nucleon. Three of them, $f_1(x,\ppSq{p})$, $g_{1L}(x,\ppSq{p})$, and $h_1(x,\ppSq{p})$ are generalizations of the three leading-twist PDFs, whereas other TMDs do not have simple collinear limits.

TMDs are able to describe a wide range of phenomena following quantum chromodynamics (QCD) factorization theorems~\cite{Rogers2016,Collins2011,Collins1985,Collins1981,Collins1982,Aybat2011,Aybat2012}. The Collins asymmetry can be explained using the transversity TMD, $h_1(x,\ppSq{p})$~\cite{Anselmino2015,Kang2016,DAlesio2020,Cammarota2020}; the double spin asymmetry $ A_{LT} $ in SIDIS can be described using the worm-gear TMD, $ g_{1T}(x,\ppSq{p}) $~\cite{Bhattacharya2021}; and one can employ the pretzelosity TMD, $ h_{1T}^{\perp}(x,\ppSq{p}) $, to describe the $ A_{UT}^{\sin(3\phi_h-\phi_S)} $ single spin asymmetry~\cite{Lefky2015a}.

The nucleon TMDs have been investigated using several QCD inspired models, e.g., MIT bag model~\cite{Avakian2010}, covariant parton model~\cite{Efremov2009,Bastami2021a}, spectator model~\cite{Bacchetta2008,Bacchetta2020a,Bacchetta2021,Bacchetta2022}, light-front quark-diquark model motivated by soft wall anti-de Sitter (AdS)/QCD~\cite{Maji2017a}, light-cone constituent model~\cite{Pasquini2008}, etc. Meanwhile, promising theoretical frameworks for accessing  TMDs also include the discretized space-time Euclidean lattice~\cite{Musch2011,Musch2012,Ji:2014hxa,Yoon2017,Constantinou:2020hdm} and the Dyson–Schwinger equations approach~\cite{Shi:2018zqd,Shi:2020pqe}. However, these approaches working in the Euclidean space-time encounter challenges in determining TMDs directly. 

In this work, we investigate the quark TMDs of the proton within basis light-front quantization (BLFQ), which provides an alternative non-perturbative framework for solving relativistic many-body bound state problems in quantum field theories~\cite{Vary2010,Maris2013,Zhao2014}. Previously, this approach has been successfully applied to explore the TMDs of the electron in QED~\cite{Hu2021}. Here we consider the light-front effective Hamiltonian for the nucleon in the constituent valence quark Fock space and solve for its mass eigenstates and light-front wavefunctions (LFWFs). Parameters in our Hamiltonian have been fixed to reproduce the nucleon mass and the flavor Dirac form factors~\cite{Xu2021,Mondal2020}. The LFWFs in this calculation have been successfully applied to compute nucleon properties such as the electromagnetic and axial form factors, radii, PDFs, GPDs, angular momentum distributions etc.~\cite{Xu2021,Mondal2020,Liu2022} Here, we extend those investigations to study the proton TMDs at the leading twist.

\section{BLFQ framework}
Basis light-front quantization (BLFQ)~\cite{Vary2010} is a non-perturbative framework for calculating the internal structures of a hadron's bound state. BLFQ starts with the light-front eigenvalue equation~\cite{Lepage1980,Pauli1985}
\begin{gather}\label{eq:schrodinger}
    H \ket{P,\Lambda}= M^2 \ket{P,\Lambda} \, ,
\end{gather}    
and adopts basis states to express it as a hermitian matrix eigenvalue problem.

The current work truncates the Fock sector expansion~\cite{Lepage1980,Brodsky2001a} of the proton system to the leading three-quark sector~\cite{Xu2021,Mondal2020}
\begin{gather}
    \scalebox{0.9}{$\begin{aligned}
        \ket{P,\Lambda}&=\sum_{\lambda_1,\lambda_2,\lambda_3} \int \frac{\prod_{i=1}^3  \rmd x_i \rmd p_i^{\perp} }{\left[2(2\pi)^3\right]^2\sqrt{x_1 x_2 x_3}} \delta(1-\sum_{i=1}^3 x_i)\times \\
        &\quad\delta^2(\sum_{i=1}^3 p_i^{\perp}) \psi^{\Lambda}_{\lambda_1,\lambda_2,\lambda_3}(p_1, p_2, p_3) \ket{\{\lambda_i,p_i\}} \, .
    \end{aligned}$}\label{eq:proton_LFWF}
\end{gather}
Here, $ P=(P^+,\frac{M^2}{P^+},0^\perp) $, $ M $ and $ \Lambda $ are the momentum, mass and light-front helicity~\cite{Soper1972} of the proton, respectively. $ p^\perp_i $ is the transverse momentum of the $ i $th quark, $ x_i=\frac{p_i^+}{P^+} $ is its longitudinal momentum fraction, $ \lambda_i $ is its light-front helicity, and roman alphabet subscripts run through the three quarks. $ \psi^{\Lambda}_{\lambda_1,\lambda_2,\lambda_3} $ is the light-front three-quark helicity amplitude.

With quarks being the only explicit degrees of freedom, the following effective Hamiltonian is diagonalized to obtain the light-front wavefunction (LFWF) of the proton state~\cite{Xu2021}
\begin{align}
    H_{\text{eff.}} &= \begin{aligned}[t]
        & \sum_{i=1}^3 \frac{m_i^2+\ppSq{p_i}}{x_i} +\frac{1}{2}\sum_{i,j=1}^3 V^{\text{conf.}}_{i,j} \\
        & +\frac{1}{2}\sum_{i,j=1}^3 V^{\text{OGE}}_{i,j} \, .
    \end{aligned}\label{eq:hamiltonian}
\end{align}
The confinement potential $  \frac{1}{2}\sum_{i,j} V^{\text{conf.}}_{i,j} $ includes both the transverse and the longitudinal confinements. The transverse confining potential is adopted from light-front holographic QCD~\cite{Brodsky2015}. We also employ a complementary longitudinal confining potential~\cite{Li2017}. The total confinement potential reduces to the 3-dimensional harmonic oscillator potential in the nonrelativistic limit~\cite{Li2016,Xu2021,Mondal2020}. The one-gluon exchange (OGE) term, a QCD version of the corresponding term in QED~\cite{Wiecki2015}, encodes the interactions among the three active quarks arising from the exchange of a gluon.

With the help of 2-dimensional harmonic oscillator (2D HO) basis states in the transverse direction\footnote{Here, $ b $ is the HO basis scale parameter, $ \theta=\arg(p^\perp) $ and $ \rho=\ppAbs{p}/b $ .}
\begin{align}\label{eq:2DHO}
    \phi_{n m}(p^{\perp})=\frac{1}{b}\sqrt{\frac{4\pi \times n!}{(n+|m|)!}}e^{im \theta}L_{n}^{|m|}(\rho^2)\rho^{|m|}e^{-\rho^2/2}\, ,
\end{align}
plane-wave state in the longitudinal direction confined in a box with length $ L $ with an anti-periodic boundary condition, and also light-cone helicity state~\cite{Soper1972} in the spin space, \eq{\ref{eq:schrodinger}} is transformed to a hermitian matrix eigenvalue problem. The above basis choice introduces four quantum numbers for every quark single-particle state: $ n,m $ for the transverse degree of freedom (d.o.f.), $ k $ for the longitudinal d.o.f. (longitudinal momentum is $ \frac{2\pi k}{L} $ with $ k $ taking half odd-integer values) and $ \lambda $ for the spin d.o.f. Two basis space truncations, $ \nmax $ and $ K $, are added to render the resulting matrix finite~\cite{Xu2021}. $ \nmax $ introduces truncation in the transverse direction for the total energy of the 2D HO basis states $ \sum_i \left( 2 n_i +|m_i| +1  \right) \leq \nmax $, and $ K $ represents the resolution in the longitudinal direction
\begin{gather}
    \sum_i k_i =K \, , \\
    x_i=\frac{p_i^+}{P^+}=\frac{k_i}{K} \, .
\end{gather}
In this paper, all the calculations are performed with $ \nmax=10, K=16.5 $. The physical parameters in the effective Hamiltonian (\eq{\ref{eq:hamiltonian}}), which include the quark mass ($ m_{\mathrm{q/k}} $) in the kinetic energy ($ \sum_{i=1}^3 \frac{m_i^2+\ppSq{p_i}}{x_i} $), the quark mass ($ m_{\mathrm{q/g}} $) and coupling constant ($ \alpha_s $) entering the OGE term ($ V^{\text{OGE}} $), and the strength ($ \kappa $) of the confinement potential ($ V^{\text{conf.}}_{i,j} $), are listed in \tab{\ref{tab:parameters}}. Along with the value of 2D HO basis scale $ b=\gev{0.6} $, those parameters are determined by fitting the nucleon mass and the flavor form factors as in \papers{\cite{Xu2021,Mondal2020}}. By fitting the same observables, the values of those parameters display decreasing changes  with increasing basis truncations, $ N_{\mathrm{max}} $ and $ K $~\cite{Xu2021}. We surmise that our TMD distribution would change slightly when varying $ N_{\mathrm{max}} $ and $ K $.

\begin{table}[htp]	
    \centering
    \begin{tabular}{cccc}
        \toprule
        $m_{\rm{q/k}}$ 	   ~&~   $m_{\rm{q/g}}$    ~&~ $\kappa$      	 ~&~ $\alpha_s$    \\
        \midrule
        $0.3$ GeV ~&~	 $0.2$	GeV    ~&~ $0.34$ GeV ~&~ $1.1\pm 0.1$ \\
        \bottomrule
    \end{tabular}
    \caption{Model parameters for the basis truncations $N_{\rm{max}}=10$ and $K=16.5$~\cite{Xu2021,Mondal2020}.}\label{tab:parameters}
\end{table}

After diagonalizing the Hamiltonian matrix, we obtain the proton mass $ M=\gev{1.018} $, and the corresponding LFWF in momentum space expressed via \eq{\ref{eq:proton_LFWF}} in terms of three-quark helicity amplitudes 

\begin{gather}
    \psi^{\Lambda}_{\lambda_1,\lambda_2,\lambda_3}(p_1,p_2,p_3)=\sum_{\{n_i,m_i\}} \psi(\alpha_1,\alpha_2,\alpha_3) \prod_{i=1}^3 \phi_{n_i m_i}(p_i^{\perp})\, .\label{eq:lfwf_proton_qqq}
\end{gather}

Here $ \alpha $ is the set of all four quantum numbers $ k,n,m,\lambda $ and $ \psi(\alpha_1,\alpha_2,\alpha_3) $ are the amplitudes of the LFWF expressed in the BLFQ basis.

\section{TMDs within the BLFQ framework}
TMDs are parameterization factors of the quark correlation function~\cite{Goeke2005,Meissner2007}
\begin{align}
    \begin{aligned}
        \Phi^{[\Gamma]}(P,&S;x=\frac{p^{\LCp}}{P^{\LCp}},p^{\LCperp})=\frac{1}{2} \int \frac{\rmd z^{\LCm}\rmd z^{\LCperp}}{2(2\pi)^3} e^{i p\cdot z} \times \\
        &\bra{P,S} \bar{\Psi}(0)\mycal{W}(0,z) \Gamma \Psi(z) \ket{P,S}|_{z^{\LCp}=0}\;,
    \end{aligned}\label{eq:TMD_vector_correlations}
\end{align}
where color and flavor indexes and summations are implicit. Here, $\Psi$ represents the quark field, and $\Gamma$ is the Dirac matrix which in the leading twist, is taken as $ \Gamma=\gamma^+,\gamma^+\gamma^5,i\sigma^{j+}\gamma^5 $. For the spin vector, we follow the same notations as in \papers{\cite{Meissner2009,Hu2021}}: 
\begin{gather}
    (S^{\LCp}, S^{\LCm}, S^{\LCperp})=(\frac{S^3 P^{\LCp}}{M}, \frac{-S^3 M}{P^{\LCp}}, S^1, S^2) \, , \\
    (S^1, S^2, S^3)=(\sin\theta \cos\varphi, \sin\theta \sin\varphi, \cos\theta) \, .
\end{gather}

In the current study, we only retain the zeroth-order expansion of the gauge link
\begin{gather}\label{eq:gauge_link}
    \mycal{W}(0,z)\approx \mybb{1}.
\end{gather}
This choice is very common in practice~\cite{Avakian2010,Maji2017a,Pasquini2008,Bastami2021a}, under which all T-odd TMDs reduce to zero.

In the leading twist, in general one would find eight TMDs, which are parameterized as follows~\cite{Goeke2005,Meissner2007}
\begin{align}
    &\Phi^{[\gamma^{\LCp}]}(x,p^{\LCperp};S)=f_{1}-\frac{\epsilon_{\LCperp}^{ij}p^{i}S^{j}}{M}f_{1T}^{\LCperp} \; , \label{eq:veccortotmdplus}\\
    &\Phi^{[\gamma^{\LCp}\gamma^{5}]}(x,p^{\LCperp};S)=S^{3}g_{1L}+\frac{p^{\LCperp}\cdot S^{\LCperp}}{M}g_{1T} \; , \label{eq:veccortotmdplus5} \\
    &\begin{aligned}
        \Phi^{[i\sigma^{j\LCp}\gamma^{5}]}&(x,p^{\LCperp};S)=S^{j}h_{1}+S^{3}\frac{p^{j}}{M}h_{1L}^{\LCperp}\\
        &+S^{i}\frac{2p^{i}p^{j}-(p^{\LCperp})^{2}\delta^{ij}}{2M^{2}}h_{1T}^{\LCperp} +\frac{\epsilon_{\LCperp}^{ji}p^{i}}{M}h_{1}^{\LCperp}\; ,\label{eq:veccortotmdsigmaj}
    \end{aligned}
\end{align}
where the convention $ \epsilon_\LCperp^{12}=1 $ is used. However, two of them, $ \fITperp{} $ and $ \hIperp{} $, are T-odd and thus vanish under the current gauge link approximation, \eq{\ref{eq:gauge_link}}.

With the help of the rotation matrix in the $ S=\frac{1}{2} $ representation of the $ SU(2) $ group~\cite{Lorce2011b,Meissner2009}, all six T-even leading-twist TMDs are expressed in terms of the light-front helicity amplitudes $ \psi^{\Lambda}_{\lambda_1,\lambda_2,\lambda_3}(p_1, p_2, p_3) $ as
\begin{align}
    f_1&=\int\rmd[123]\sum_{\lambda_2\lambda_3}\left[|\psi^{+}_{+\lambda_2\lambda_3}|^2+|\psi^{+}_{-\lambda_2\lambda_3}|^2\right],\label{eq:f1}\\
    g_{1L}&=\int\rmd[123]\sum_{\lambda_2\lambda_3}\left[|\psi^{+}_{+\lambda_2\lambda_3}|^2-|\psi^{+}_{-\lambda_2\lambda_3}|^2\right],\label{eq:g1L}\\
    g_{1T}&=\frac{2M}{\ppSq{p}}\int\rmd[123]\sum_{\lambda_2\lambda_3}\mathfrak{Re}\left[p^R \psi^{+\, *}_{+\lambda_2\lambda_3} \psi^{-}_{+\lambda_2\lambda_3} \right], \label{eq:g1T}\\
    h_1&=\int\rmd[123]\sum_{\lambda_2\lambda_3}\mathfrak{Re}\left[\psi^{+\, *}_{+\lambda_2\lambda_3}\psi^{-}_{-\lambda_2\lambda_3}\right],\label{eq:h1}\\
    \hILperp{}&=\frac{2M}{p^\perp}\int\rmd[123]\sum_{\lambda_2\lambda_3}\mathfrak{Re}\left[p^R \psi^{+\, *}_{-\lambda_2\lambda_3}\psi^{+}_{+\lambda_2\lambda_3}\right], \label{eq:h1Lperp}\\
    \hITperp{}&=\frac{2M^2}{(p^\perp)^4}\int\rmd[123]\sum_{\lambda_2\lambda_3}\left[(p^R)^2 \psi^{+\, *}_{-\lambda_2\lambda_3}\psi^{-}_{+\lambda_2\lambda_3}\right] .\label{eq:h1Tperp}
\end{align}
Here
\begin{align}
    \begin{aligned}
        \rmd[123]=&\frac{1}{[2(2\pi)^3]^2}\prod_{i=1}^3 (\rmd x_i \rmd^2 p_i^\perp) \delta(1-\sum_{i=1}^3 x_i) \times\\
        &\delta^2(\sum_{i=1}^3 p_i^\perp) \delta(x_1-x) \delta^2(p_1^\perp-p^\perp)
    \end{aligned}\, ,
\end{align}
$ p^R=p^1+ip^2 $, and we omit the arguments $ (x,\ppSq{p}) $ for quantities on the left and $ (p_1,p_2,p_3) $ for quantities on the right.

\begin{figure*}
    \centering
    \includegraphics[width=0.46\textwidth]{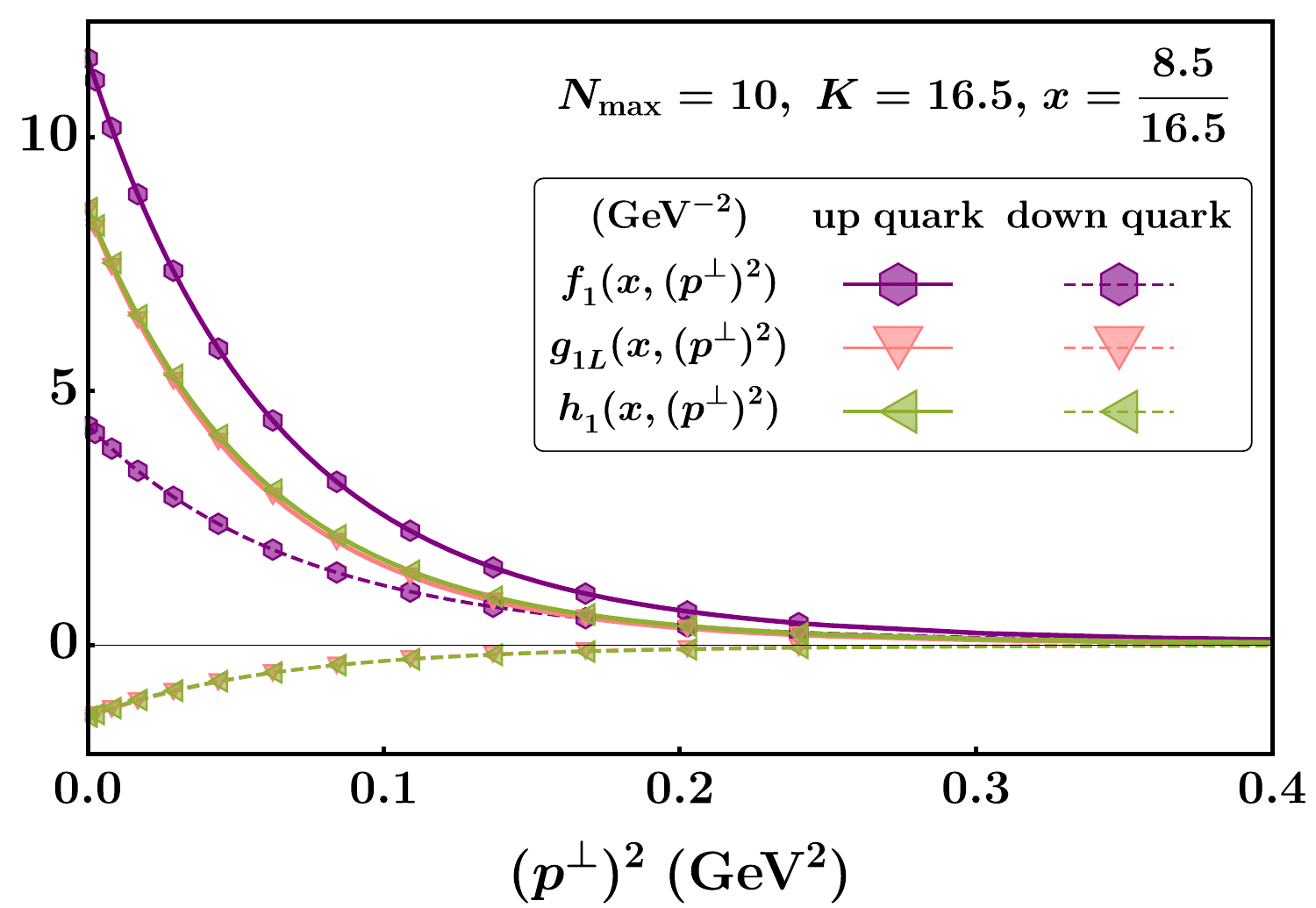}
    \includegraphics[width=0.46\textwidth]{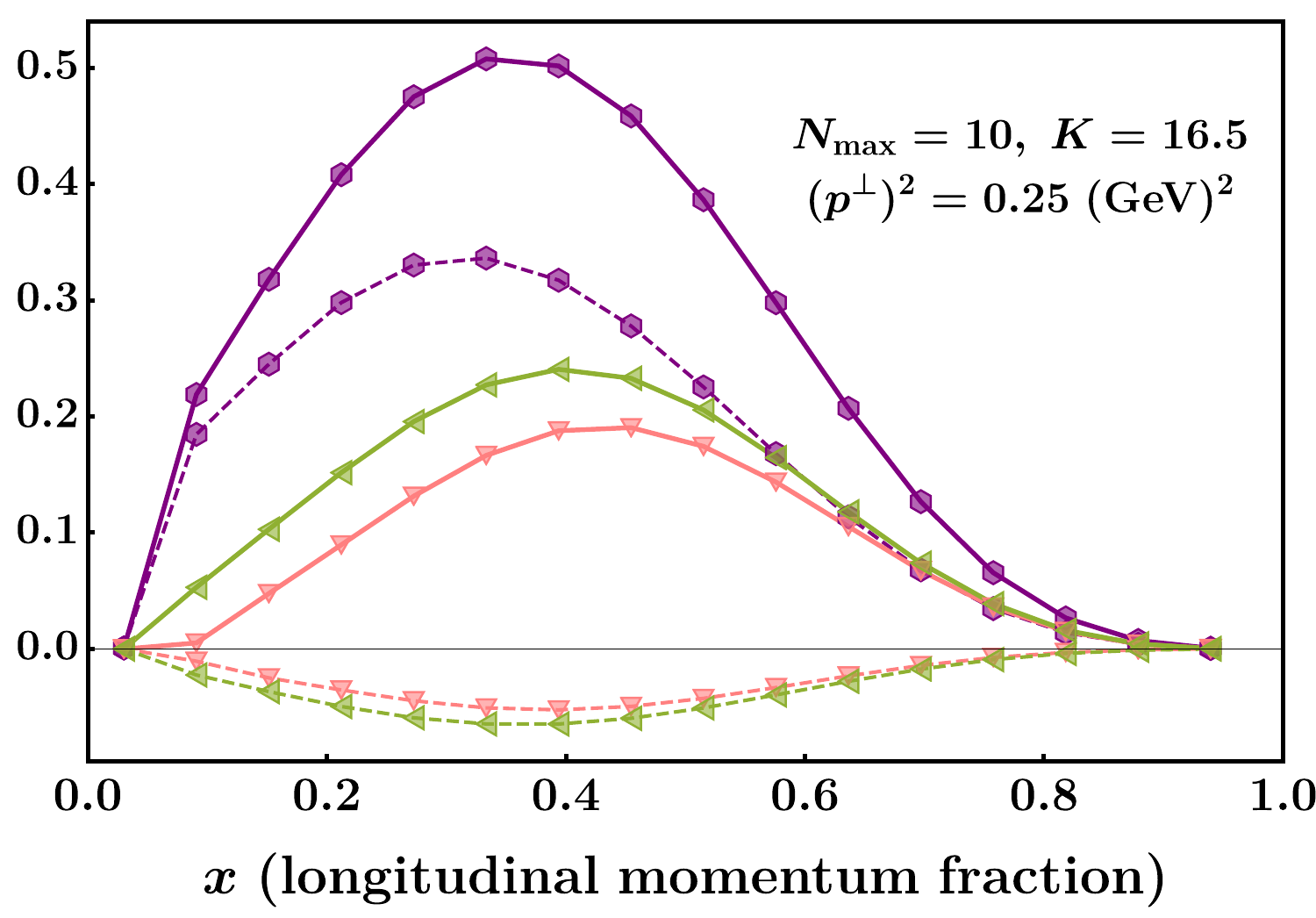}
    \includegraphics[width=0.46\textwidth]{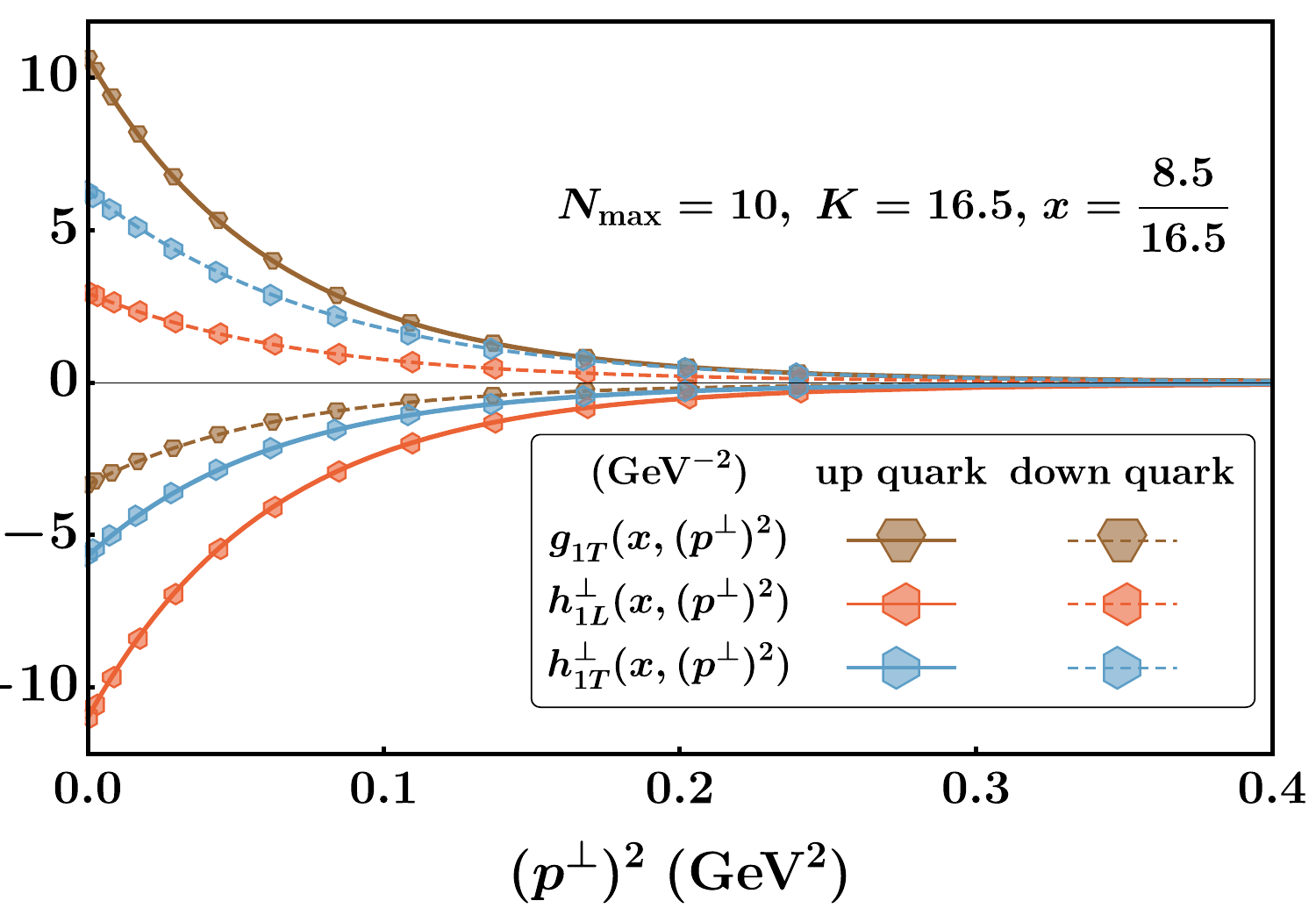}
    \includegraphics[width=0.46\textwidth]{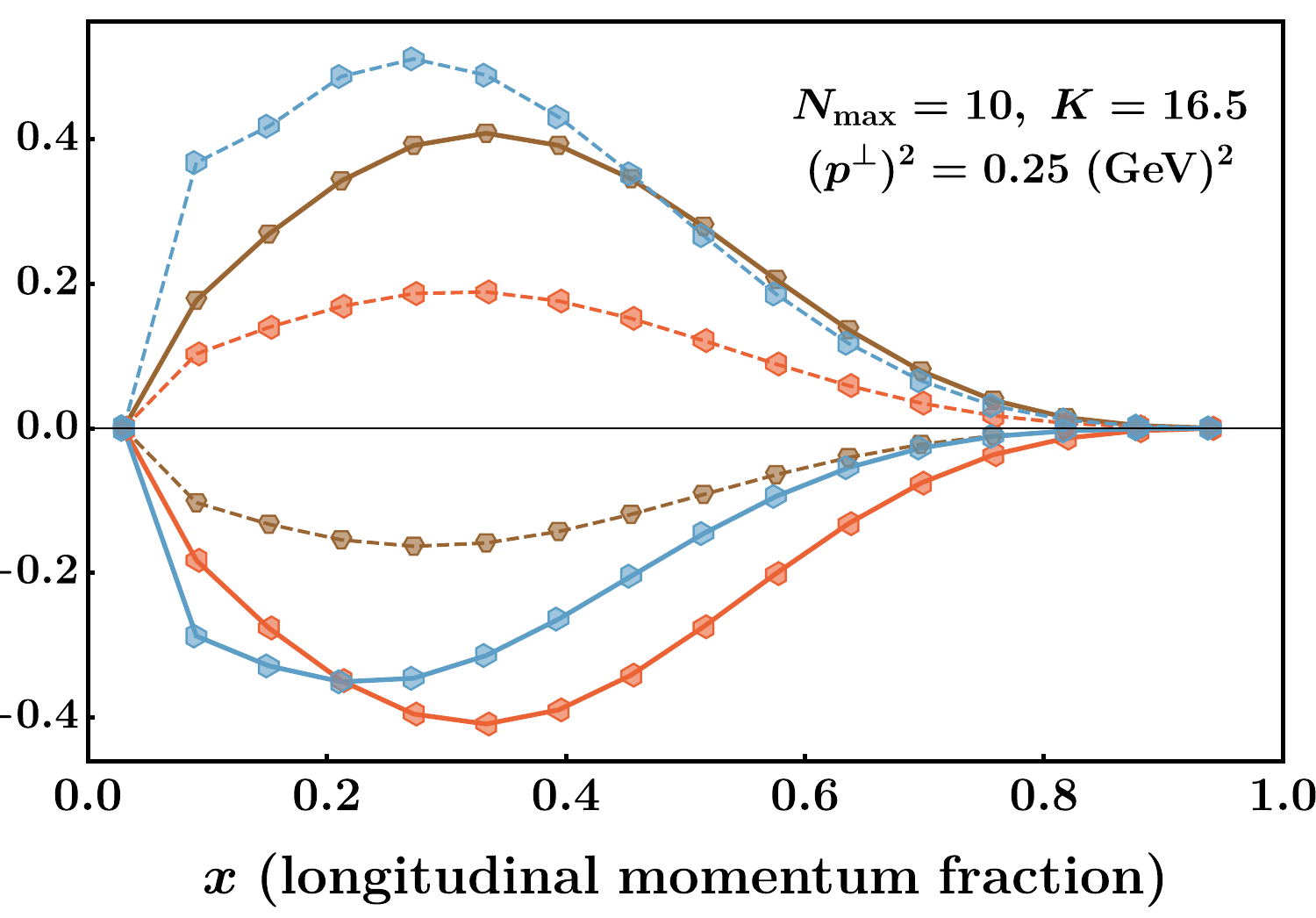}
    \caption{\label{fig:BLFQ_2D_1}(Color online) BLFQ results in the transverse direction at $ x=\frac{8.5}{16.5} $ (left), and in the longitudinal direction at $  \ppSq{p}=\gevSq{0.25}  $ (right). Plots in the upper (lower) panels are TMDs which have (do not have) proper PDF limits. Lines with different markers (colors) represent different TMDs as indicated in the legends. Solid lines represent $ u $ quark distributions and dashed lines represent $ d $ quark distributions.}
\end{figure*}

In \fig{\ref{fig:BLFQ_2D_1}}, we show our results for those TMDs in the transverse and longitudinal directions separately. We note that the qualitative behaviors of our TMDs are similar to those of other theoretical calculations in \papers{\cite{Avakian2010,Efremov2009,Bastami2021a,Bacchetta2008,Maji2017a,Pasquini2008,Musch2011,Musch2012,Ji:2014hxa,Yoon2017,Constantinou:2020hdm,Shi:2018zqd,Shi:2020pqe}}.

\subsection{Soffer-type bounds} 
Since the current calculations subsume the gluon dynamics into effective interactions among the 3 valence quarks and we ignore the gluon contributions from the gauge link, we cannot directly access the dynamical role of the gluons. These choices result in many interesting relations connecting twist-2 and twist-3 TMDs~\cite{Bacchetta2007,Goeke2003,Mulders1995}. But, due to the focus of this paper on twist-2 TMDs, we will defer the study of the validity of those relations to a future work.

Still, in \paper{\cite{Bacchetta2000}}, the authors investigated the bounds of the leading-twist TMDs from the point of view of the positivity of the matrix representing the quark helicity structure. In our current calculations, we obtain zero $ \fITp{} $ and $ \hIperp{} $ under the approximation \eq{\ref{eq:gauge_link}}. Thus, the bounds of \paper{\cite{Bacchetta2000}} reduce to
\begin{align}
    \vert h_1 \vert &\le  
    \frac{1}{2}\left( f_1 + g_{1L}\right)\,,\label{eq:bound1}\\
    \frac{(p^\perp)^2}{M^2}\vert h_{1T}^{\perp}\vert &\le 
    \left( f_1 - g_{1L}\right)\,,\label{eq:bound2}\\
    \frac{(p^\perp)^2}{M^2}\left(\gIT{}\right)^2 &\le 
    \left( f_1 + g_{1L}\right)\left( f_1 - g_{1L}\right)\, ,\label{eq:bound3}\\
    \frac{(p^\perp)^2}{M^2}\left(\hILperp{}\right)^2 &\le 
    \left( f_1 + g_{1L}\right)\left( f_1 - g_{1L}\right)\, .\label{eq:bound4}
\end{align}
We identify the points where the left-hand side (LHS) quantities and the right-right side (RHS) quantities of those four bounds above are nearest to each other and list them in {\tab{\ref{tab:bounds}}}. We find that the BLFQ results fulfill all the above bounds, which serves as an important consistency check of the current BLFQ calculations of T-even TMDs.

\begin{table*}
    \centering\renewcommand{\arraystretch}{1.2}
    \begin{tabular}{l c c c c}
        \toprule
         & LHS ($ \mathrm{GeV^{-2}} $) & RHS ($ \mathrm{GeV^{-2}} $) & $\ppSq{p}$ ($ \gevSq{} $) & $x$ \\\midrule
        \eq{\mbox{\ref{eq:bound1}}} for up quark & $2.37$ & $ 2.75 $ & $ 0.0001 $ & $ 11.5/16.5 $ \\
        \eq{\mbox{\ref{eq:bound1}}} for down quark & $ 5.261\times 10^{-3} $ & $ 5.262\times 10^{-3} $ & $0.0001$ & $ 15.5/16.5 $\\\midrule
        \eq{\mbox{\ref{eq:bound2}}} for up quark & $1.12\times 10^{-5}$ & $ 2.84\times 10^{-5} $ & 0.3025 & $ 0.5/16.5 $ \\
        \eq{\mbox{\ref{eq:bound2}}} for down quark & $ 1.60\times 10^{-5} $ & $ 1.91\times 10^{-5} $ & $ 0.3025 $ & $ 0.5/16.5 $ \\\midrule
        \eq{\mbox{\ref{eq:bound3}}} for up quark & $ 0.318 $ & $ 1.65 $ & $ 0.1089 $ & $ 1.5/16.5 $ \\
        \eq{\mbox{\ref{eq:bound3}}} for down quark & $ 0.203 $ & $ 1.66 $ & $ 0.0729 $ & $ 0.5/16.5 $ \\\midrule
        \eq{\mbox{\ref{eq:bound4}}} for up quark & $ 0.611 $ & $ 2.75 $ & $0.068$ & $0.5/16.5$ \\
        \eq{\mbox{\ref{eq:bound4}}} for down quark & $0.112\times 10^{-5}$ & $0.123\times 10^{-4}$ & $0.81$ & $ 10.5/16.5 $ \\\bottomrule
    \end{tabular}
    \caption{To investigate the Soffer-type bounds, \eqs{\ref{eq:bound1}-\ref{eq:bound4}}, we identify the points where the left-hand side (LHS) quantities and the right-right side (RHS) quantities of those four bounds are nearest to each other. We list the left-hand side quantities, right-hand side quantities and the corresponding  $ \ppSq{p} $ and $ x $ values.\label{tab:bounds}}
\end{table*}

\subsection{Reduction to the collinear distributions}
\begin{figure}
    \centering
    \includegraphics[width=0.46\textwidth]{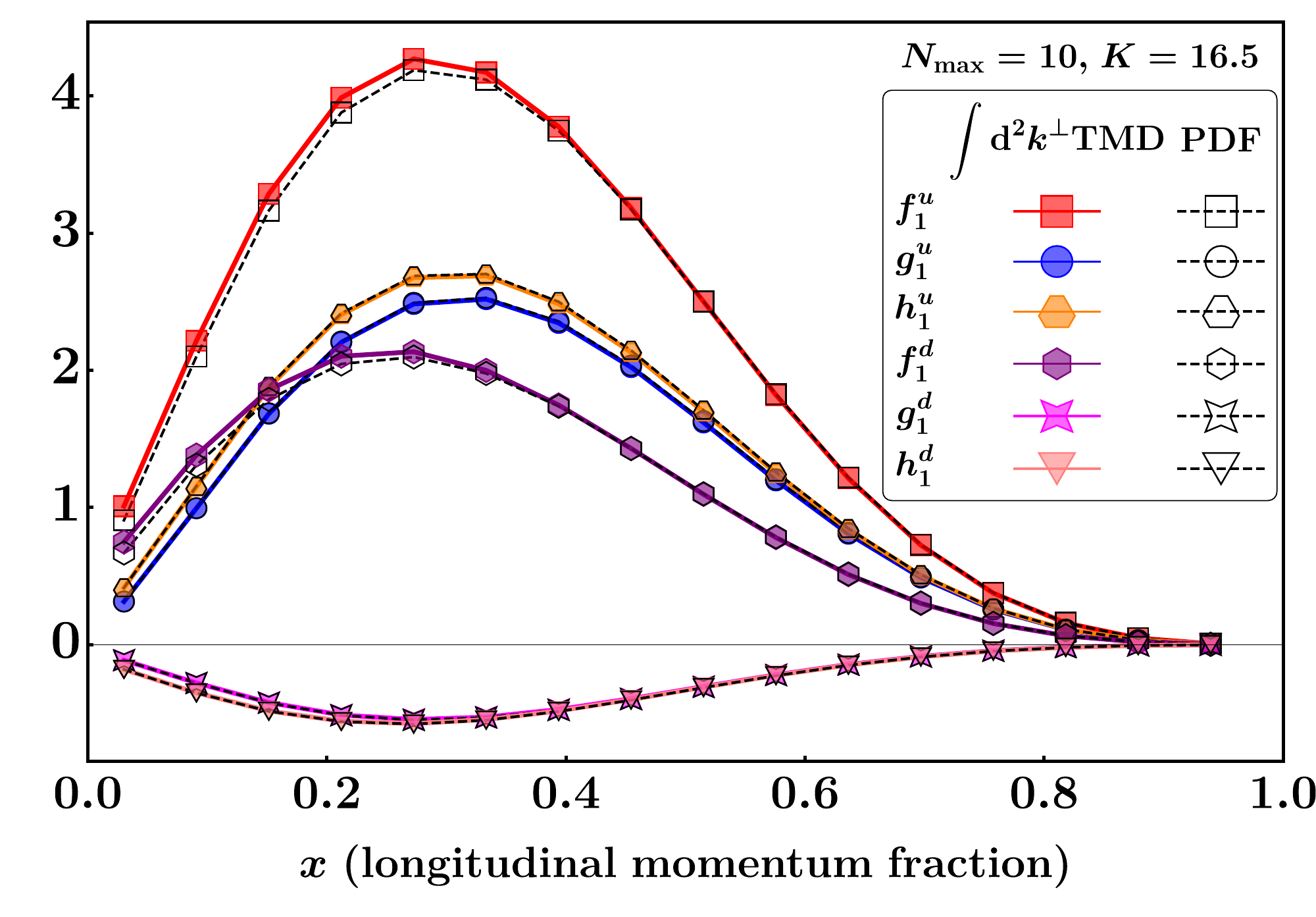}
    \caption{(Color online) Comparisons between the leading-twist integrated TMDs and PDFs~\cite{Xu2021,Mondal2020}, both calculated within the BLFQ framework. Lines with different markers represent different TMDs/PFDs as indicated in the legends. Solid colored lines represent integration over the transverse momentum of TMDs and dashed black lines represent the corresponding PDFs.\label{fig:PDFcomparison}}
\end{figure}

TMDs are the extension of collinear parton distributions (PDFs) that incorporate information in the transverse momentum direction. After integrating over the transverse momenta one should regain PDFs from TMDs. Out of the $ 8 $ leading-twist TMDs, only three of them survive after this integration. In \fig{\ref{fig:PDFcomparison}}, we plot the integration of TMDs and the corresponding PDFs calculated directly within the BLFQ framework. One observes that they compare well with each other to within small residual differences that provide metrics for our numerical uncertainties.

Further, \papers{\cite{Xu2021,Mondal2020}} evolve the same leading-twist PDFs calculated within the BLFQ framework via DGLAP equations, and find good consistency between the evolved BLFQ results and experimental results. This suggests that our current TMD calculations may have the potential to explain experimental data, a research area for a future investigation.

\subsection{Flavor-ratio results compared with the lattice QCD calculations}

\begin{figure}
    \includegraphics[width=0.46\textwidth]{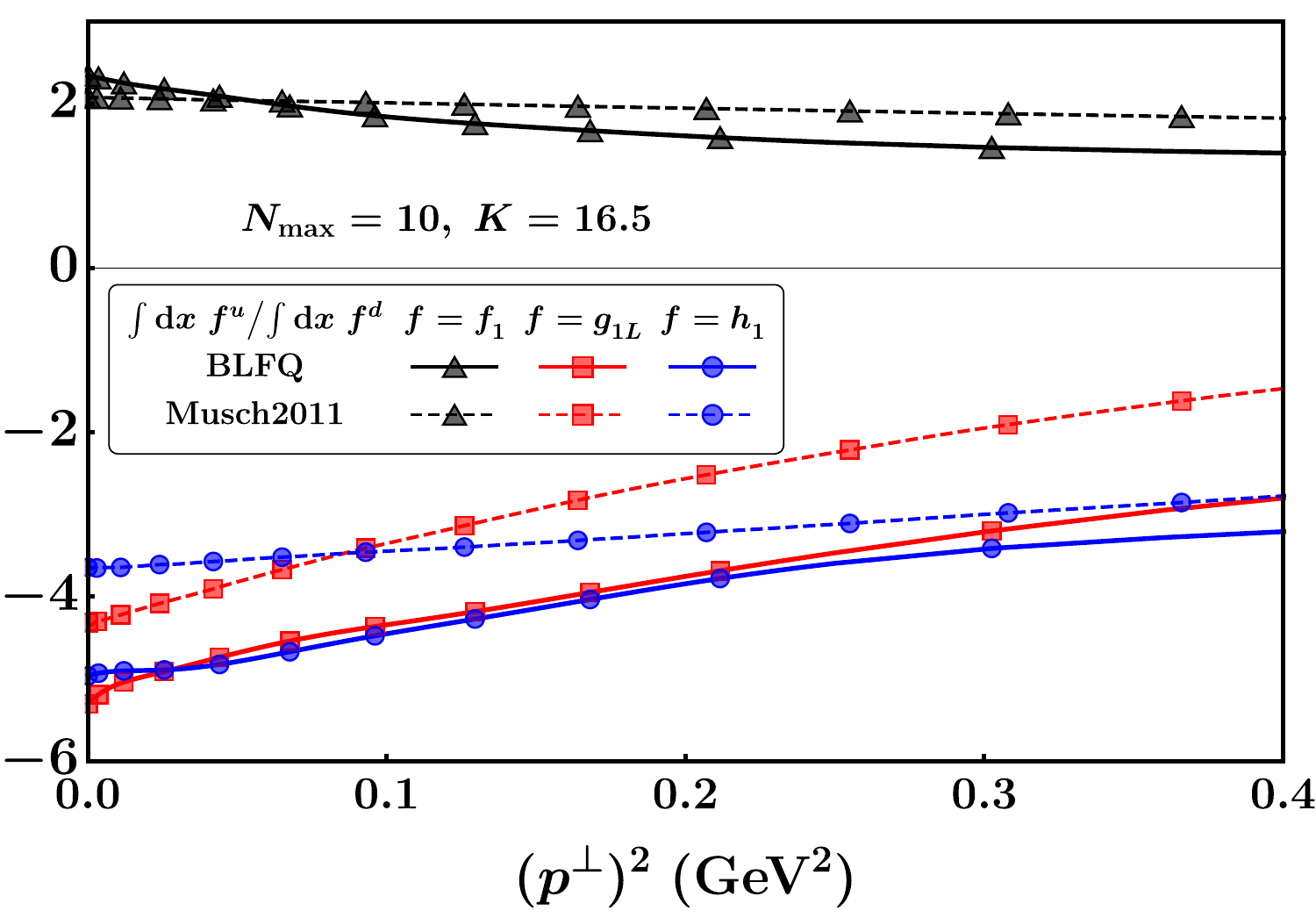}
    \caption{\label{fig:xintegration_falourratio_comparison_f1g1Lh1}(Color online) Comparisons of $ \left. \int\rmd x f^u \middle/ \int\rmd x f^d \right. $ for $ \fI{} $, $ \gIL{} $ and $ \hI{} $ from the BLFQ framework and the central values from the lattice QCD simulations obtained by parameterizing $ \widetilde{A}_2$, $\widetilde{A}_6 $ and $ \widetilde{A}_{9m} $~\cite{Musch2011}. Lines with different markers (colors) represent flavor-ratios of different distributions as indicated in the legend. Solid lines represent the BLFQ results and dashed lines represent lattice QCD results.}
\end{figure}

Reference \cite{Musch2011} calculated TMDs using lattice QCD with the assumption of a straight-line gauge link. This non-trivial gauge link also leads to vanishing T-odd TMDs like our approximation \eq{\ref{eq:gauge_link}}. In addition to the bare results, \paper{\cite{Musch2011}} also shows their results in the form of flavor-ratios $ \frac{\int \rmd x f^u}{\int \rmd x f^d} $. We adopt this quantity for cross-comparison, since it cancels, at least some of, the possible model-dependent overall factors and even scale evolution effects\footnote{From \papers{\cite{Scimemi2020,Kang2016}}, one may find that the equations and parameters for scale evolution are both flavor-independent.}. We show these comparisons in \fig{\ref{fig:xintegration_falourratio_comparison_f1g1Lh1}}.

Surprisingly, even though those results are obtained within two totally different frameworks, we still find qualitative agreement while interesting differences are visible. For example, we generally find that the magnitudes of our flavor-ratios decrease faster in the high $ \ppSq{p} $ region. We attribute this difference to the fact that, the BLFQ results for $ d $ quark are generally wider than those of the $ u $ quark in the transverse momentum (see also \fig{\ref{fig:average_pperpsquareI}} and the surrounding discussions), while in the lattice QCD simulations, they tend towards similar widths (see \figs{12, 13} of \paper{\cite{Musch2011}}).

\section{Discussions}\label{ss:discussion}

\subsection{$ x-p^{\perp} $ factorization and Gaussian ansatz}
Many preliminary extractions of TMDs from experimental data, like \paper{\cite{Anselmino2014}} for $ \fI{}$, \paper{\cite{Bhattacharya2021}} for $\gIT{}$, \papers{\cite{Anselmino2015,DAlesio2020,Cammarota2020}} for $\hI{} $ and \paper{\cite{Lefky2015a}} for $ \hITperp{} $, follow a simple functional form (the so-called Gaussian ansatz):
\begin{gather}
    f^q(x,(p^\perp)^2) = f^q(x) \frac{e^{-\frac{(p^\perp)^2}{\braket{(p^\perp)^2}_f}}}{\pi \braket{(p^\perp)^2}_f} \, .
\end{gather}
Here, $f^q(x,(p^\perp)^2)$ is a generic notation of all TMDs for flavor $ q $ and $ f^q(x) $ is its collinear part, \ie{}, $ f^q(x)=\int \rmd^2 p^\perp f^q(x,(p^\perp)^2)  $. Apart from the specific Gaussian distribution, the most important implication of the above ansatz is that the averaged transverse momentum squared $ \braket{(p^\perp)^2} $ is distribution-dependent, but flavor and $ x $ independent.

But, more realistic extractions, like \papers{\cite{Bacchetta2017,Bertone2019,Signori2013,Scimemi2020}} for $ \fI{} $ and \paper{\cite{Kang2016}} for $ \hI{} $, do not support those simplifications. Thus, it would be very interesting to investigate whether those assumptions hold for the BLFQ results, which may facilitate future comparisons with the experimental extractions and may even guide future extractions.

\subsubsection{Flavor and $ x $ dependence of $\braket{(p^\perp)^2}$ within the BLFQ framework}
\begin{figure*}
    \centering
    \includegraphics[width=0.44\textwidth]{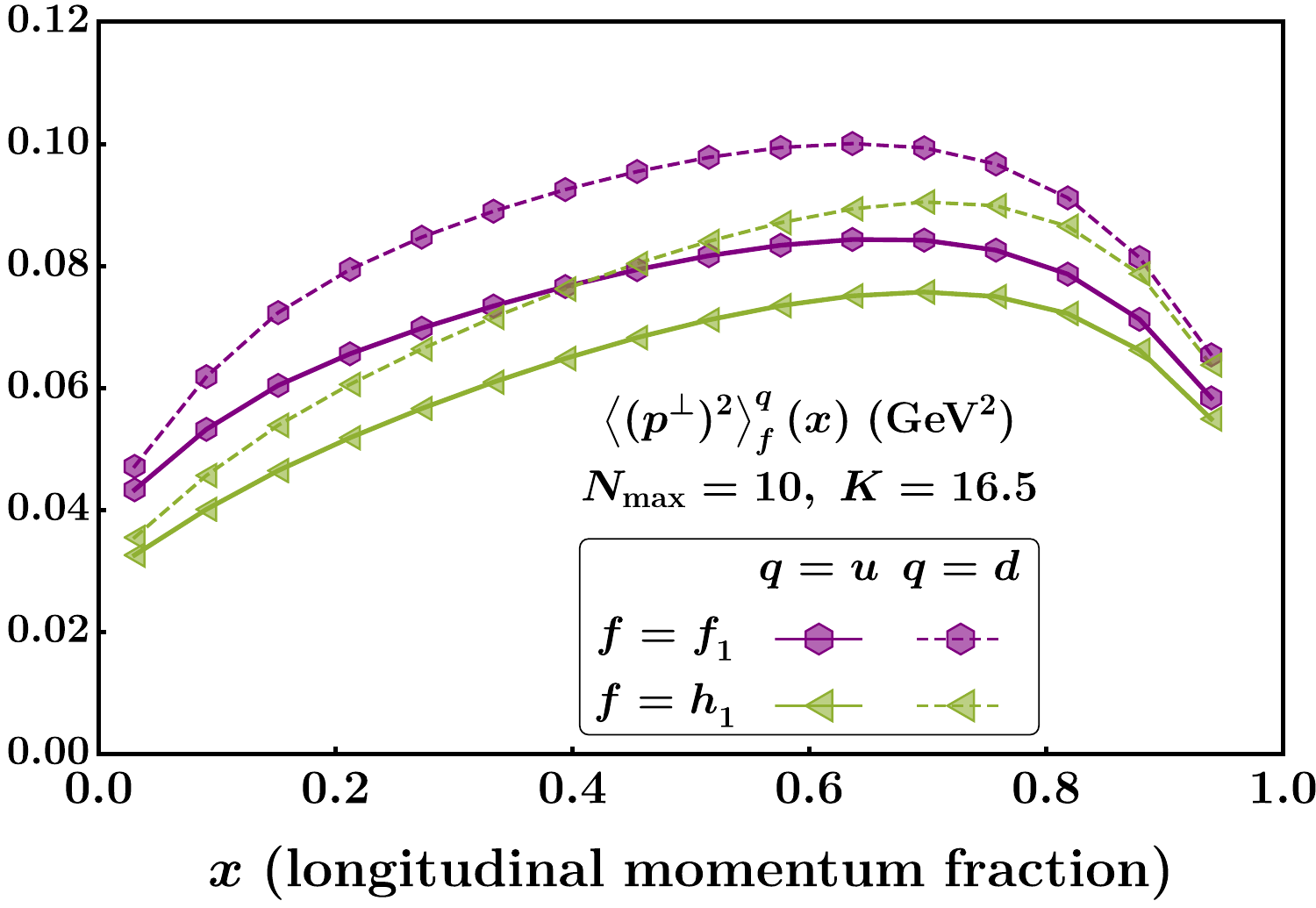}
    \includegraphics[width=0.44\textwidth]{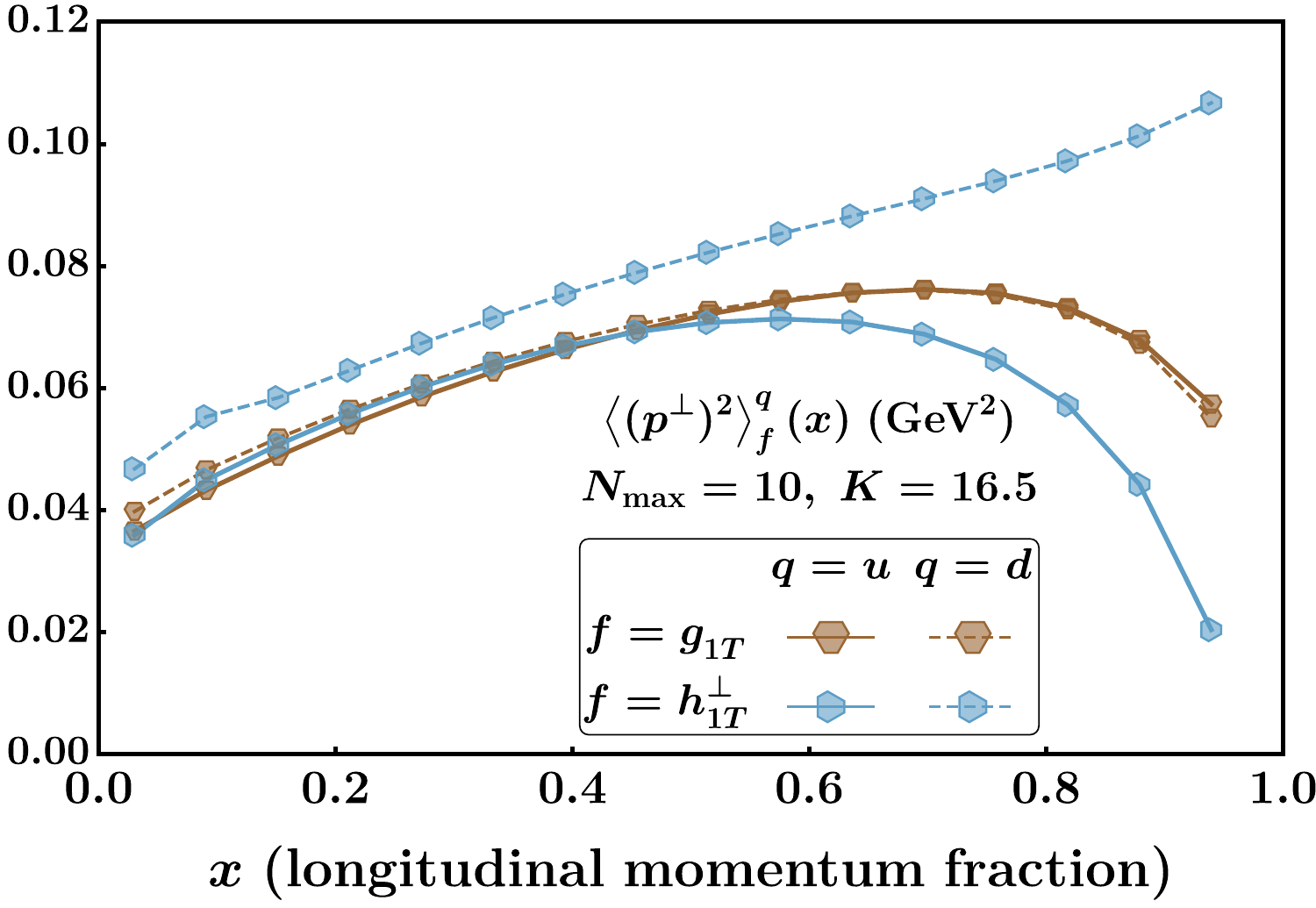}
    \caption{(Color online) $ x $ and flavor dependence of $ \braket{(p^\perp)^2}_f^q(x) $ (\eq{\ref{eq:pperpSqI}}) of the BLFQ results for $ \fI{}, \gIT{}, \hI{}$ and $\hITperp{} $. Lines with different markers (colors) represent $ \braket{(p^\perp)^2}_f^q(x) $ obtained from different TMDs as indicated in the legends. Solid lines represent calculations of $ u $ quark and dashed lines $ d $ quark. \label{fig:average_pperpsquareI}}
\end{figure*}

We compute the averaged transverse momentum squared $ \braket{\ppSq{p}} $ for the BLFQ results as
\begin{align}
    \braket{(p^\perp)^2}_f^q(x) = \frac{\int\rmd^2 p^\perp\; (p^\perp)^2 f_{\text{BLFQ}}^q(x,(p^\perp)^2)}{\int\rmd^2 p^\perp \; f_{\text{BLFQ}}^q(x,(p^\perp)^2)} \, ,
\end{align}
where $ f_{\text{BLFQ}}^q $ are TMDs obtained within the BLFQ framework. The averaged transverse momentum squared for $ \fI{}, \gIT{}, \hI{}$ and $\hITperp{} $ are shown in \fig{\ref{fig:average_pperpsquareI}}. One can see that within the BLFQ framework, $ \braket{(p^\perp)^2} $ do exhibit a strong flavor and $ x $ dependence. It is also observed that $ \braket{(p^\perp)^2} $ for $ d $ quarks is generally larger than that of $ u $ quark.

We further fit the $ x $ dependence of $ \braket{(p^\perp)^2}_f^q $ for different TMDs and flavors using the following function
\begin{gather}
    \braket{(p^\perp)^2}_f^q(x)=a_f^q x^{b_f^q} + c_f^q x^{d_f^q} +e_f^q \, . \label{eq:pperpSqI_fitting}
\end{gather}
We find that, excluding $ \hITperp{d} $, the $ x $ dependences of $ \braket{(p^\perp)^2}_f^q $ from all other TMDs are generally very uniform, with parameters very close to the following average values (unit of all the dimension-2 values are $ \gevSq{} $):
\begin{gather}
    \braket{(p^\perp)^2}_f^q(x)\approx 0.08 x^{0.49} -0.06 x^{6.62} +0.02 \label{eq:pperpSqI_parameter}\, .
\end{gather}

Qualitatively, the above results are consistent with those experimental extractions which do take into account the $ x $-dependence of $ \braket{\ppSq{p}} $, like \papers{\cite{Signori2013,Bacchetta2017}}. The preliminary results, \eqs{\ref{eq:pperpSqI_fitting}, \ref{eq:pperpSqI_parameter}}, are useful as an alternative for the functional form of the averaged transverse momentum squared for future experimental extractions.

\subsubsection{Gaussian ansatz and the BLFQ results}
We then investigate the compatibility of the BLFQ results with the Gaussian ansatz. For this purpose, we use two methods to `fit' the Gaussian width. As the first method, we calculate the $ x $-independent $ \braket{\ppSq{p}} $ commonly used in the literature as:
\begin{gather}
    \left.\braket{(p^\perp)^2}_f^q\right|_{\text{I}} = \frac{\int\rmd x \rmd^2 p^\perp\; (p^\perp)^2 f_{\text{BLFQ}}^q(x,(p^\perp)^2)}{\int\rmd x \rmd^2 p^\perp \; f_{\text{BLFQ}}^q(x,(p^\perp)^2)} \, .\label{eq:pperpSqI}
\end{gather}
As the second method, we determine the $ x $-dependent $ \braket{\ppSq{p}} $ by demanding that the Gaussian distributions coincide with the BLFQ results at $ \ppSq{p}=0 $\footnote{For TMDs $ \gIT{},\hILperp{},\hITperp{} $ (\eq{\ref{eq:g1T}, \ref{eq:h1Lperp}, \ref{eq:h1Tperp}}), we actually use a small ($ \mycal{O}(10^{-4}) $) non-zero $ \ppSq{p} $, since in the numerical calculation, we are not able to take the denominator to be zero.}. This is a commonly used strategy, like in \paper{\cite{Avakian2010}}, to investigate the compatibility between Gaussian ansatz and TMD calculations. We have:
\begin{gather}
    \left.\braket{(p^\perp)^2}_f^q\right|_{\text{II}} =  \frac{f^q_{\text{BLFQ}}(x)}{\pi f^q_{\text{BLFQ}}(x,0)} \, .\label{eq:pperpSqII}
\end{gather}

Using those two Gaussian widths we construct two different Gaussian-type distributions as
\begin{gather}
    f_{\text{Gaus.}}^q(x,(p^\perp)^2) = f_{\text{BLFQ}}^q(x) \frac{e^{-\frac{(p^\perp)^2}{\braket{(p^\perp)^2}_f^q}}}{\pi \braket{(p^\perp)^2}_f^q}  \, , \label{eq:BLFQGaussian}
\end{gather}
and compare them with the BLFQ results in the small ($ \lesssim \gevSq{0.4}$) $ \ppSq{p} $ region in \fig{\ref{fig:gaussian_small}}. From these comparisons, it is evident that in the small $ \ppSq{p} $ region, if, and only if we include proper $ x $-dependence of the Gaussian width, then the Gaussian distribution would be a good approximation for the BLFQ results.

\newlength{\plotlength}
\setlength{\plotlength}{0.46\textwidth}
\begin{figure}
    \centering
    \includegraphics[width=\plotlength]{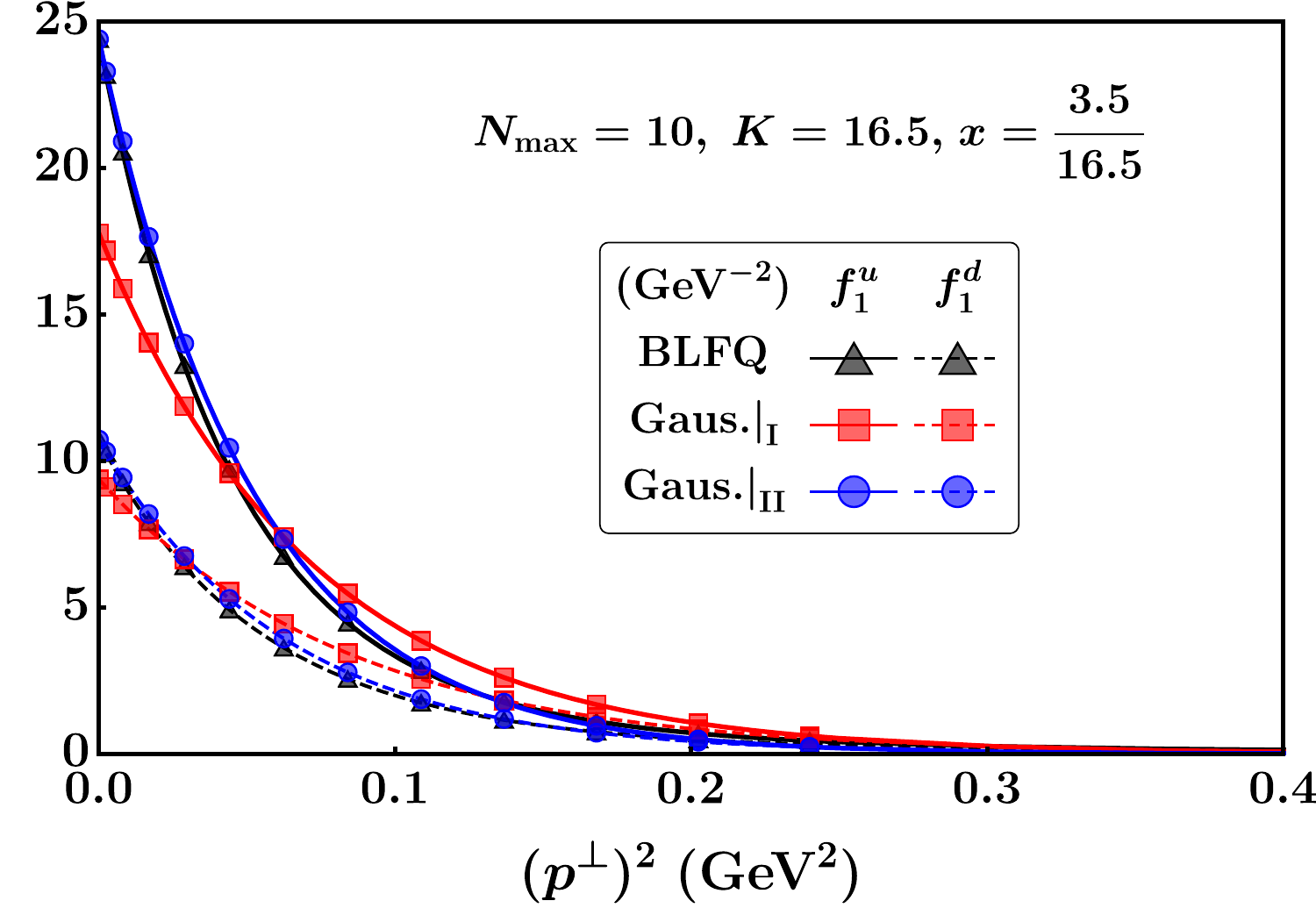}
    \includegraphics[width=\plotlength]{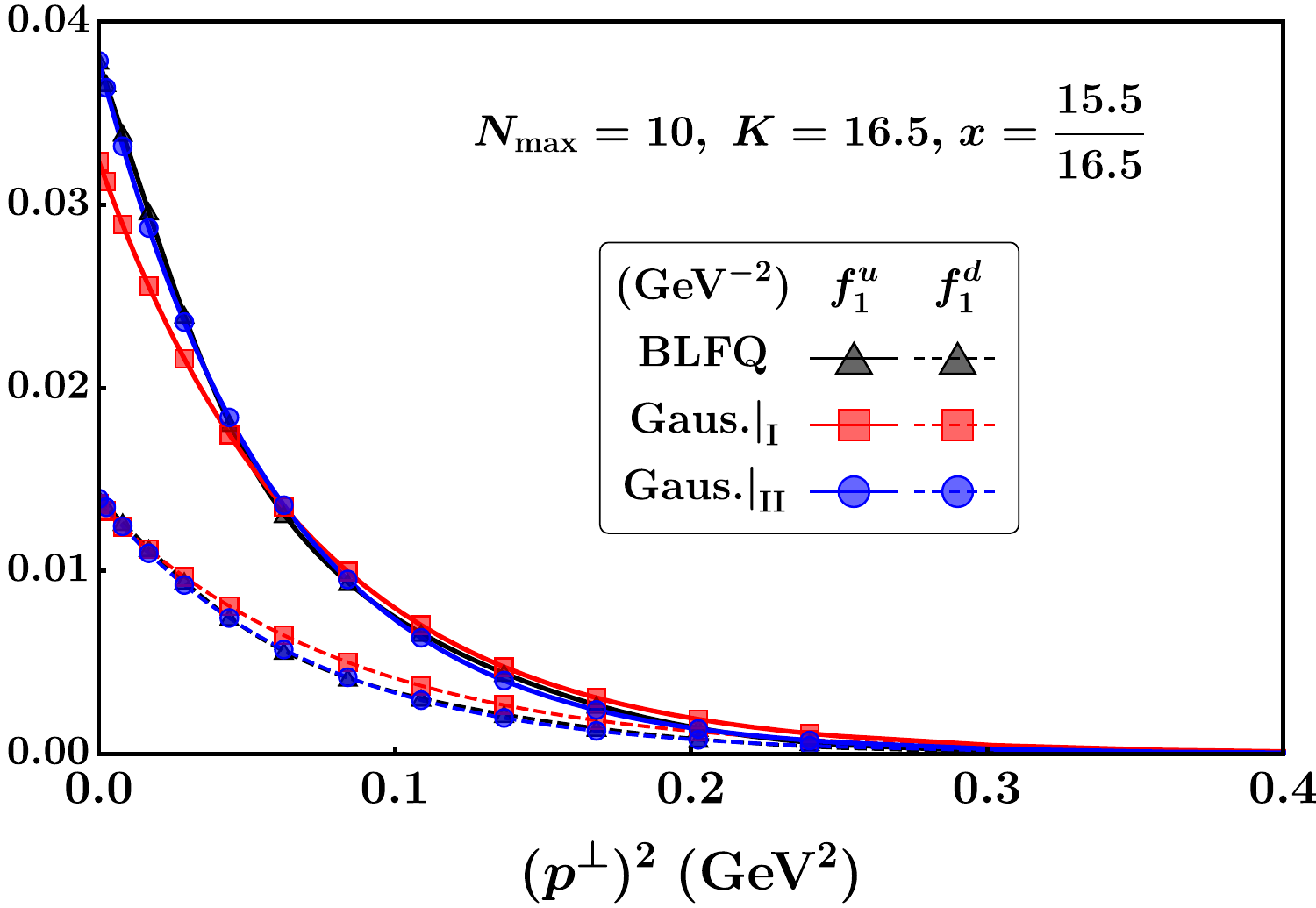}
    \caption{(Color online) Comparisons between the bare BLFQ results and Gaussian-type distributions (\eq{\ref{eq:BLFQGaussian}}) with Gaussian widths obtained from different methods. ``$ \left.\text{Gaus.}\right|_\text{I} $'' (``$ \left.\text{Gaus.}\right|_\text{II} $'') use the Gaussian width from \eq{\ref{eq:pperpSqI}} (\eq{\ref{eq:pperpSqII}}). We show comparisons for $ \fI{} $ in small ($ \lesssim \gevSq{0.4} $) $ \ppSq{p} $ region at $ x=\frac{3.5}{16.5} $ ($ x=\frac{15.5}{16.5} $) in the upper (lower) panel. Lines with different markers (colors) represent distributions from different methods as indicated in the legends. Solid lines represent $ u $ quark distributions and dashed lines represent $ d $ quark distributions. \label{fig:gaussian_small}}
\end{figure}

\begin{figure}
    \centering
    \includegraphics[width=\plotlength]{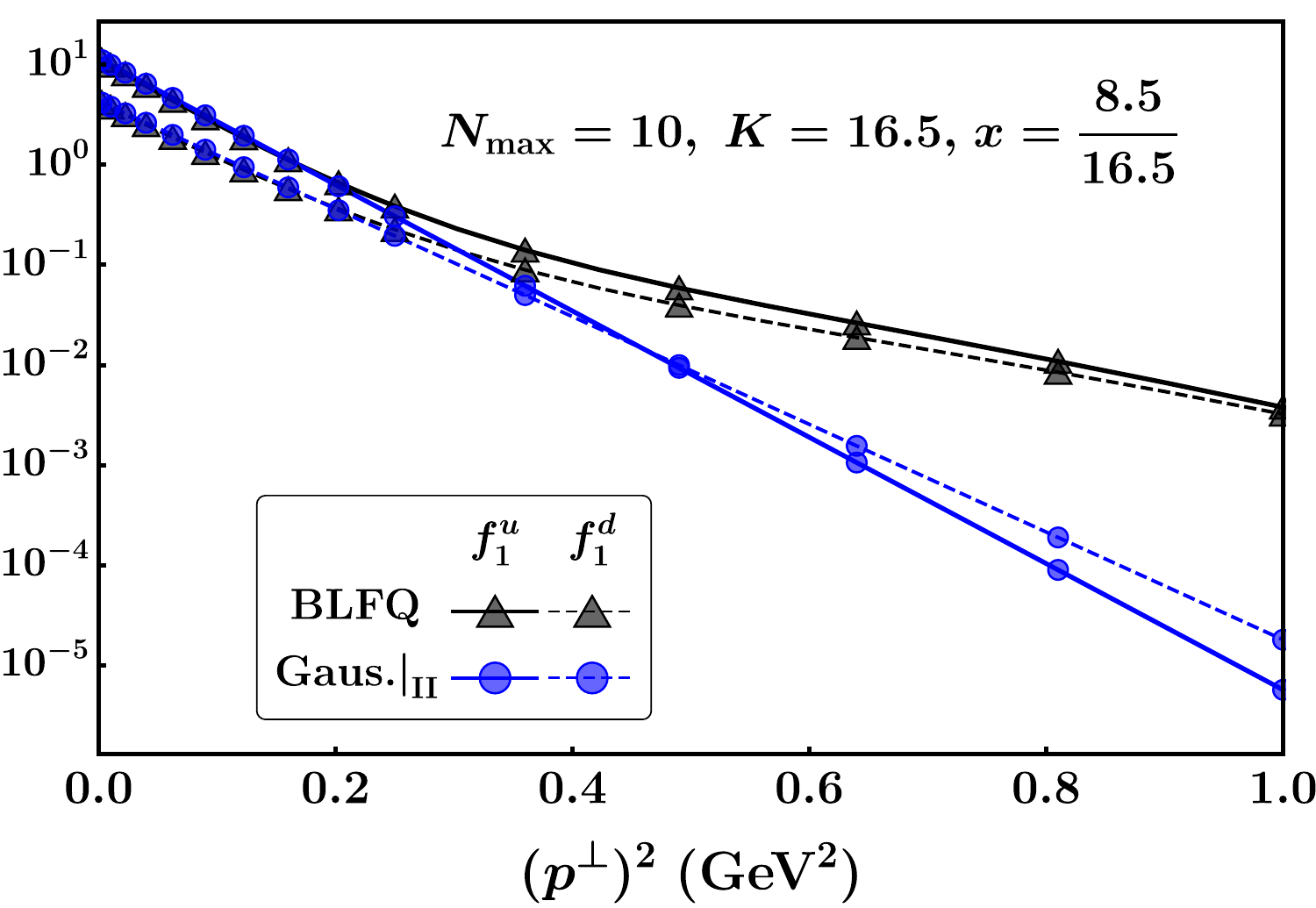}
    \caption{(Color online) Comparisons between the BLFQ results and the Gaussian-type distributions (\eqs{\ref{eq:pperpSqII}, \ref{eq:BLFQGaussian}}) for $ \fI{} $ in the linear-log plot at $ x=\frac{8.5}{16.5} $. Lines with different markers (colors) represent distributions from different methods as indicated in the legend. Solid lines represent $ u $ quark distributions and dashed lines represent $ d $ quark distributions. \label{fig:gaussian_big}}
\end{figure}

In \fig{\ref{fig:gaussian_big}}, we show the comparisons between the BLFQ results and the Gaussian-type distributions in the linear-log plot to investigate their large momentum behaviors. Since the large momentum behaviors of the Gaussian-type distributions are very similar to each other, here we only compare with the Gaussian type results obtained from \eqs{\ref{eq:pperpSqII}, \ref{eq:BLFQGaussian}}. From these comparisons, it is evident that the BLFQ results decrease more slowly than the Gaussian-type distributions. This is expected, since in the large $ \ppSq{p} $ region, TMDs would decrease as an inverse power of $ p^\perp $~\cite{Bacchetta2008a}, which we believe is reasonably approximated within BLFQ's dynamics.

Due to the space limits, we only show the plots of $ \fI{} $ for the above two comparisons, but the observations are similar for other five leading-twist T-even TMDs.

\subsection{Model-dependent relations}

In full QCD, all TMDs should be independent of each other. But, some non-trivial relations between TMDs are also observed in many quark models.\footnote{See \papers{\cite{Avakian2010,Lorce2011b}} and references therein.} We find that none of those previously found relations are satisfied by the BLFQ results, suggesting that the current BLFQ results of the leading-twist T-even TMDs may indeed be independent of each other. 

We surmise that having independent T-even TMDs provides support for our underlying non-perturbative framework. The absence of model-dependent relations, along with the fact that our results follow the universal Soffer-type bounds, implies that the BLFQ framework are heading in a valuable direction for simulating full QCD.

\section{Conclusions}
Basis Light-front Quantization (BLFQ) has been proposed as a non-perturbative framework for solving quantum field theory. In this work, we have calculated the quark TMD PDFs for the proton from its light-front wave functions within the framework of BLFQ. These wave functions have been obtained from the eigenvectors of an effective light-front Hamiltonian in the leading Fock sector incorporating a three-dimensional confining potential and a one-gluon exchange interaction with fixed coupling.

In this study, the gauge link has been set to unity, which leaves us six nonzero TMDs (T-even) out of the eight leading-twist TMDs. We compare our results with the previous PDF calculations within the same framework and with the lattice QCD simulation, and find good consistency in both cases. The validity of the universal Soffer-type inequalities and the absence of all the previously found model-dependent relations together imply that the BLFQ framework captures key elements of the non-perturbative physics from QCD. Increasing the number of Fock sectors would generate more independent helicity amplitudes, and thus more independent TMDs from higher twist or from the T-odd domain. One would then expect that extensions to higher Fock sectors would bring us closer to our ultimate goal, the description of full QCD.

Our calculations do not support the $ x-p^\perp $ factorization commonly used in the preliminary phenomenological studies~\cite{Anselmino2014,Bhattacharya2021,Anselmino2015,DAlesio2020,Cammarota2020,Lefky2015a}. More specifically, the non-trivial $ x $ dependence of $ \braket{(p^\perp)^2}_f^q $ precludes the $ x-p^\perp $ factorization of the form
\begin{gather}
    f^q(x,\ppSq{p}) = f^q(x)\frac{\hat{f^q}(\ppSq{p})}{\mycal{N}} \, ,
\end{gather}
where $ \mycal{N}=\int \rmd^2 p^\perp \hat{f^q}(\ppSq{p}) $. We also compare the BLFQ results with Gaussian-type distributions and find that Gaussian distribution is only useful for describing the BLFQ results in the small $ \ppSq{p} $ region.

Future developments will focus on the inclusion of a non-trivial gauge link that will provide a prediction of the Boer-Mulders and the Sivers functions and their application to spin-asymmetries. Another major development will focus on the extension to higher Fock sectors, especially the $ \ket{qqqq\bar{q}} $ and $ \ket{qqqg} $ Fock sectors, to evaluate gluon and sea-quark TMDs. Our approach can also be utilized to calculate higher-twist TMDs.

\section*{Acknowledgements}                                                 
C. M. is supported by new faculty start up funding by the Institute of Modern Physics, Chinese Academy of Sciences, Grant No. E129952YR0. C. M. also thanks the Chinese Academy of Sciences Presidents International Fellowship Initiative for the support via Grants No. 2021PM0023. X. Z. is supported by new faculty startup funding by the Institute of Modern Physics, Chinese Academy of Sciences, by Key Research Program of Frontier Sciences, Chinese Academy of Sciences, Grant No. ZDB-SLY-7020, by the Natural Science Foundation of Gansu Province, China, Grant No. 20JR10RA067 by the Foundation for Key Talents of Gansu Province, by the Central Funds Guiding the Local Science and Technology Development of Gansu Province, and by the Strategic Priority Research Program of the Chinese Academy of Sciences, Grant No. XDB34000000. J. P. V. is supported by the Department of Energy under Grants No. DE-FG02-87ER40371, and No. DE-SC0018223 (SciDAC4/NUCLEI). This research use resources of the National Energy Research Scientific Computing Center (NERSC), a U.S. Department of Energy Office of Science User Facility operated under Contract No. DE-AC02-05CH11231. A portion of the computational resources were also provided by Gansu Computing Center.

\biboptions{sort&compress}
\bibliographystyle{elsarticle-num}
\bibliography{hep-ph.bib}

\begin{thebibliography}{10}
\expandafter\ifx\csname url\endcsname\relax
  \def\url#1{\texttt{#1}}\fi
\expandafter\ifx\csname urlprefix\endcsname\relax\def\urlprefix{URL }\fi
\expandafter\ifx\csname href\endcsname\relax
  \def\href#1#2{#2} \def\path#1{#1}\fi

\bibitem{ZEUS:2003pwh}
S.~Chekanov, et~al., {Measurement of deeply virtual Compton scattering at
  HERA}, Phys. Lett. B 573 (2003) 46--62.
\newblock \href {http://arxiv.org/abs/hep-ex/0305028}
  {\path{arXiv:hep-ex/0305028}}, \href
  {http://dx.doi.org/10.1016/j.physletb.2003.08.048}
  {\path{doi:10.1016/j.physletb.2003.08.048}}.

\bibitem{COMPASS:2008isr}
M.~Alekseev, et~al., {Collins and Sivers asymmetries for pions and kaons in
  muon-deuteron DIS}, Phys. Lett. B 673 (2009) 127--135.
\newblock \href {http://arxiv.org/abs/0802.2160} {\path{arXiv:0802.2160}},
  \href {http://dx.doi.org/10.1016/j.physletb.2009.01.060}
  {\path{doi:10.1016/j.physletb.2009.01.060}}.

\bibitem{HERMES:2009lmz}
A.~Airapetian, et~al., {Observation of the Naive-T-odd Sivers Effect in
  Deep-Inelastic Scattering}, Phys. Rev. Lett. 103 (2009) 152002.
\newblock \href {http://arxiv.org/abs/0906.3918} {\path{arXiv:0906.3918}},
  \href {http://dx.doi.org/10.1103/PhysRevLett.103.152002}
  {\path{doi:10.1103/PhysRevLett.103.152002}}.

\bibitem{Collins1985}
J.~C. Collins, D.~E. Soper, G.~Sterman, {Transverse momentum distribution in
  Drell-Yan pair and W and Z boson production}, Nuclear Physics, Section B
  250~(1-4) (1985) 199--224.
\newblock \href {http://dx.doi.org/10.1016/0550-3213(85)90479-1}
  {\path{doi:10.1016/0550-3213(85)90479-1}}.

\bibitem{Collins2011}
J.~C. Collins,
  \href{http://ebooks.cambridge.org/ref/id/CBO9780511975592}{{Foundations of
  Perturbative QCD}}, Cambridge University Press, Cambridge, 2011.
\newblock \href {http://dx.doi.org/10.1017/CBO9780511975592}
  {\path{doi:10.1017/CBO9780511975592}}.
\newline\urlprefix\url{http://ebooks.cambridge.org/ref/id/CBO9780511975592}

\bibitem{Bacchetta2007}
A.~Bacchetta, M.~Diehl, K.~Goeke, A.~Metz, P.~J. Mulders, M.~Schlegel,
  {Semi-inclusive deep inelastic scattering at small transverse momentum},
  Journal of High Energy Physics 2007~(2).
\newblock \href {http://arxiv.org/abs/0611265} {\path{arXiv:0611265}}, \href
  {http://dx.doi.org/10.1088/1126-6708/2007/02/093}
  {\path{doi:10.1088/1126-6708/2007/02/093}}.

\bibitem{Ji2005}
X.~Ji, J.~P. Ma, F.~Yuan, {QCD factorization for semi-inclusive deep-inelastic
  scattering at low transverse momentum}, Physical Review D - Particles,
  Fields, Gravitation and Cosmology 71~(3) (2005) 1--19.
\newblock \href {http://arxiv.org/abs/0404183} {\path{arXiv:0404183}}, \href
  {http://dx.doi.org/10.1103/PhysRevD.71.034005}
  {\path{doi:10.1103/PhysRevD.71.034005}}.

\bibitem{Tangerman:1994eh}
R.~D. Tangerman, P.~J. Mulders, {Intrinsic transverse momentum and the
  polarized Drell-Yan process}, Phys. Rev. D 51 (1995) 3357--3372.
\newblock \href {http://arxiv.org/abs/hep-ph/9403227}
  {\path{arXiv:hep-ph/9403227}}, \href
  {http://dx.doi.org/10.1103/PhysRevD.51.3357}
  {\path{doi:10.1103/PhysRevD.51.3357}}.

\bibitem{Collins:2002kn}
J.~C. Collins, {Leading twist single transverse-spin asymmetries: Drell-Yan and
  deep inelastic scattering}, Phys. Lett. B 536 (2002) 43--48.
\newblock \href {http://arxiv.org/abs/hep-ph/0204004}
  {\path{arXiv:hep-ph/0204004}}, \href
  {http://dx.doi.org/10.1016/S0370-2693(02)01819-1}
  {\path{doi:10.1016/S0370-2693(02)01819-1}}.

\bibitem{Zhou:2009jm}
J.~Zhou, F.~Yuan, Z.-T. Liang, {Transverse momentum dependent quark
  distributions and polarized Drell-Yan processes}, Phys. Rev. D 81 (2010)
  054008.
\newblock \href {http://arxiv.org/abs/0909.2238} {\path{arXiv:0909.2238}},
  \href {http://dx.doi.org/10.1103/PhysRevD.81.054008}
  {\path{doi:10.1103/PhysRevD.81.054008}}.

\bibitem{Diehl2003}
M.~Diehl, {Generalized parton distributions}, Physics Reports 388~(2-4) (2003)
  41--277.
\newblock \href {http://arxiv.org/abs/0307382} {\path{arXiv:0307382}}, \href
  {http://dx.doi.org/10.1016/j.physrep.2003.08.002}
  {\path{doi:10.1016/j.physrep.2003.08.002}}.

\bibitem{Belitsky:2005qn}
A.~V. Belitsky, A.~V. Radyushkin, {Unraveling hadron structure with generalized
  parton distributions}, Phys. Rept. 418 (2005) 1--387.
\newblock \href {http://arxiv.org/abs/hep-ph/0504030}
  {\path{arXiv:hep-ph/0504030}}, \href
  {http://dx.doi.org/10.1016/j.physrep.2005.06.002}
  {\path{doi:10.1016/j.physrep.2005.06.002}}.

\bibitem{Ji:1996nm}
X.-D. Ji, {Deeply virtual Compton scattering}, Phys. Rev. D 55 (1997)
  7114--7125.
\newblock \href {http://arxiv.org/abs/hep-ph/9609381}
  {\path{arXiv:hep-ph/9609381}}, \href
  {http://dx.doi.org/10.1103/PhysRevD.55.7114}
  {\path{doi:10.1103/PhysRevD.55.7114}}.

\bibitem{Goeke:2001tz}
K.~Goeke, M.~V. Polyakov, M.~Vanderhaeghen, {Hard exclusive reactions and the
  structure of hadrons}, Prog. Part. Nucl. Phys. 47 (2001) 401--515.
\newblock \href {http://arxiv.org/abs/hep-ph/0106012}
  {\path{arXiv:hep-ph/0106012}}, \href
  {http://dx.doi.org/10.1016/S0146-6410(01)00158-2}
  {\path{doi:10.1016/S0146-6410(01)00158-2}}.

\bibitem{Goloskokov2008}
S.~V. Goloskokov, P.~Kroll, {The role of the quark and gluon GPDs in hard
  vector-meson electroproduction}, European Physical Journal C 53~(3) (2008)
  367--384.
\newblock \href {http://dx.doi.org/10.1140/epjc/s10052-007-0466-5}
  {\path{doi:10.1140/epjc/s10052-007-0466-5}}.

\bibitem{Collins1997}
J.~C. Collins, L.~Frankfurt, M.~Strikman, {Factorization for hard exclusive
  electroproduction of mesons in QCD}, Physical Review D - Particles, Fields,
  Gravitation and Cosmology 56~(5) (1997) 2982--3006.
\newblock \href {http://arxiv.org/abs/9611433} {\path{arXiv:9611433}}, \href
  {http://dx.doi.org/10.1103/PhysRevD.56.2982}
  {\path{doi:10.1103/PhysRevD.56.2982}}.

\bibitem{Rogers2016}
T.~C. Rogers, {An overview of transverse-momentum-dependent factorization and
  evolution}, European Physical Journal A 52~(6) (2016) 1--12.
\newblock \href {http://arxiv.org/abs/1509.04766} {\path{arXiv:1509.04766}},
  \href {http://dx.doi.org/10.1140/epja/i2016-16153-7}
  {\path{doi:10.1140/epja/i2016-16153-7}}.

\bibitem{Collins1981}
J.~C. Collins, D.~E. Soper, {Back-to-back jets in QCD}, Nuclear Physics,
  Section B 193~(2) (1981) 381--443.
\newblock \href {http://dx.doi.org/10.1016/0550-3213(81)90339-4}
  {\path{doi:10.1016/0550-3213(81)90339-4}}.

\bibitem{Collins1982}
J.~C. Collins, D.~E. Soper, {Parton distribution and decay functions}, Nuclear
  Physics, Section B 194~(3) (1982) 445--492.
\newblock \href {http://dx.doi.org/10.1016/0550-3213(82)90021-9}
  {\path{doi:10.1016/0550-3213(82)90021-9}}.

\bibitem{Aybat2011}
S.~M. Aybat, T.~C. Rogers,
  \href{https://link.aps.org/doi/10.1103/PhysRevD.83.114042}{{Transverse
  momentum dependent parton distribution and fragmentation functions with QCD
  evolution}}, Physical Review D - Particles, Fields, Gravitation and Cosmology
  83~(11) (2011) 114042.
\newblock \href {http://dx.doi.org/10.1103/PhysRevD.83.114042}
  {\path{doi:10.1103/PhysRevD.83.114042}}.
\newline\urlprefix\url{https://link.aps.org/doi/10.1103/PhysRevD.83.114042}

\bibitem{Aybat2012}
S.~M. Aybat, J.~C. Collins, J.~W. Qiu, T.~C. Rogers,
  \href{https://link.aps.org/doi/10.1103/PhysRevD.85.034043}{{QCD evolution of
  the Sivers function}}, Physical Review D - Particles, Fields, Gravitation and
  Cosmology 85~(3) (2012) 034043.
\newblock \href {http://arxiv.org/abs/1110.6428} {\path{arXiv:1110.6428}},
  \href {http://dx.doi.org/10.1103/PhysRevD.85.034043}
  {\path{doi:10.1103/PhysRevD.85.034043}}.
\newline\urlprefix\url{https://link.aps.org/doi/10.1103/PhysRevD.85.034043}

\bibitem{Anselmino2015}
M.~Anselmino, M.~Boglione, U.~D'Alesio, J.~O. Hernandez, S.~Melis, F.~Murgia,
  A.~Prokudin, {Collins functions for pions from SIDIS and new e+e- data: A
  first glance at their transverse momentum dependence}, Physical Review D -
  Particles, Fields, Gravitation and Cosmology 92~(11).
\newblock \href {http://dx.doi.org/10.1103/PhysRevD.92.114023}
  {\path{doi:10.1103/PhysRevD.92.114023}}.

\bibitem{Kang2016}
Z.~B. Kang, A.~Prokudin, P.~Sun, F.~Yuan, {Extraction of quark transversity
  distribution and Collins fragmentation functions with QCD evolution},
  Physical Review D 93~(1).
\newblock \href {http://arxiv.org/abs/1505.05589} {\path{arXiv:1505.05589}},
  \href {http://dx.doi.org/10.1103/PhysRevD.93.014009}
  {\path{doi:10.1103/PhysRevD.93.014009}}.

\bibitem{DAlesio2020}
U.~D'Alesio, C.~Flore, A.~Prokudin,
  \href{https://doi.org/10.1016/j.physletb.2020.135347}{{Role of the Soffer
  bound in determination of transversity and the tensor charge}}, Physics
  Letters, Section B: Nuclear, Elementary Particle and High-Energy Physics 803
  (2020) 135347.
\newblock \href {http://dx.doi.org/10.1016/j.physletb.2020.135347}
  {\path{doi:10.1016/j.physletb.2020.135347}}.
\newline\urlprefix\url{https://doi.org/10.1016/j.physletb.2020.135347}

\bibitem{Cammarota2020}
J.~Cammarota, L.~P. Gamberg, Z.~B. Kang, J.~A. Miller, D.~Pitonyak,
  A.~Prokudin, T.~C. Rogers, N.~Sato,
  \href{https://doi.org/10.1103/PhysRevD.102.054002}{{Origin of single
  transverse-spin asymmetries in high-energy collisions}}, Physical Review D
  102~(5) (2020) 54002.
\newblock \href {http://arxiv.org/abs/2002.08384} {\path{arXiv:2002.08384}},
  \href {http://dx.doi.org/10.1103/PhysRevD.102.054002}
  {\path{doi:10.1103/PhysRevD.102.054002}}.
\newline\urlprefix\url{https://doi.org/10.1103/PhysRevD.102.054002}

\bibitem{Bhattacharya2021}
S.~Bhattacharya, Z.-B. Kang, A.~Metz, G.~Penn, D.~Pitonyak,
  \href{http://arxiv.org/abs/2110.10253}{{First global QCD analysis of the TMD
  g1T from semi-inclusive DIS data}}, Not published yet\href
  {http://arxiv.org/abs/2110.10253} {\path{arXiv:2110.10253}}.
\newline\urlprefix\url{http://arxiv.org/abs/2110.10253}

\bibitem{Lefky2015a}
C.~Lefky, A.~Prokudin, {Extraction of the distribution function h1T from
  experimental data}, Physical Review D - Particles, Fields, Gravitation and
  Cosmology 91~(3) (2015) 1--14.
\newblock \href {http://arxiv.org/abs/1411.0580} {\path{arXiv:1411.0580}},
  \href {http://dx.doi.org/10.1103/PhysRevD.91.034010}
  {\path{doi:10.1103/PhysRevD.91.034010}}.

\bibitem{Avakian2010}
H.~Avakian, A.~V. Efremov, P.~Schweitzer, F.~Yuan, {Transverse momentum
  dependent distribution functions in the bag model}, Physical Review D -
  Particles, Fields, Gravitation and Cosmology 81~(7) (2010) 1--21.
\newblock \href {http://dx.doi.org/10.1103/PhysRevD.81.074035}
  {\path{doi:10.1103/PhysRevD.81.074035}}.

\bibitem{Efremov2009}
A.~V. Efremov, P.~Schweitzer, O.~V. Teryaev, P.~Zavada, {Transverse momentum
  dependent distribution functions in a covariant parton model approach with
  quark orbital motion}, Physical Review D - Particles, Fields, Gravitation and
  Cosmology 80~(1) (2009) 1--13.
\newblock \href {http://dx.doi.org/10.1103/PhysRevD.80.014021}
  {\path{doi:10.1103/PhysRevD.80.014021}}.

\bibitem{Bastami2021a}
S.~Bastami, A.~V. Efremov, P.~Schweitzer, O.~V. Teryaev, P.~Zavada,
  \href{https://doi.org/10.1103/PhysRevD.103.014024}{{Structure of the nucleon
  at leading and subleading twist in the covariant parton model}}, Physical
  Review D 103~(1) (2021) 14024.
\newblock \href {http://dx.doi.org/10.1103/PhysRevD.103.014024}
  {\path{doi:10.1103/PhysRevD.103.014024}}.
\newline\urlprefix\url{https://doi.org/10.1103/PhysRevD.103.014024}

\bibitem{Bacchetta2008}
A.~Bacchetta, F.~Conti, M.~Radici,
  \href{https://link.aps.org/doi/10.1103/PhysRevD.78.074010}{{Transverse-momentum
  distributions in a diquark spectator model}}, Physical Review D 78~(7) (2008)
  074010.
\newblock \href {http://dx.doi.org/10.1103/PhysRevD.78.074010}
  {\path{doi:10.1103/PhysRevD.78.074010}}.
\newline\urlprefix\url{https://link.aps.org/doi/10.1103/PhysRevD.78.074010}

\bibitem{Bacchetta2020a}
A.~Bacchetta, F.~G. Celiberto, M.~Radici, P.~Taels,
  \href{https://doi.org/10.1140/epjc/s10052-020-8327-6}{{Transverse-momentum-dependent
  gluon distribution functions in a spectator model}}, European Physical
  Journal C 80~(8) (2020) 1--11.
\newblock \href {http://arxiv.org/abs/2005.02288} {\path{arXiv:2005.02288}},
  \href {http://dx.doi.org/10.1140/epjc/s10052-020-8327-6}
  {\path{doi:10.1140/epjc/s10052-020-8327-6}}.
\newline\urlprefix\url{https://doi.org/10.1140/epjc/s10052-020-8327-6}

\bibitem{Bacchetta2021}
A.~Bacchetta, F.~G. Celiberto, M.~Radici, P.~Taels,
  \href{http://arxiv.org/abs/2107.13446}{{A spectator-model way to
  transverse-momentum-dependent gluon distribution functions}} (2021) 1--7\href
  {http://arxiv.org/abs/2107.13446} {\path{arXiv:2107.13446}}.
\newline\urlprefix\url{http://arxiv.org/abs/2107.13446}

\bibitem{Bacchetta2022}
A.~Bacchetta, F.~G. Celiberto, M.~Radici,
  \href{https://pos.sissa.it/398/376}{{Toward twist-2 {\$}T{\$}-odd
  transverse-momentum-dependent gluon distributions: the {\$}f{\$}-type Sivers
  function}}, in: Proceedings of The European Physical Society Conference on
  High Energy Physics — PoS(EPS-HEP2021), Sissa Medialab, Trieste, Italy,
  2022, p. 376.
\newblock \href {http://arxiv.org/abs/arXiv:2201.10508v1}
  {\path{arXiv:arXiv:2201.10508v1}}, \href
  {http://dx.doi.org/10.22323/1.398.0376} {\path{doi:10.22323/1.398.0376}}.
\newline\urlprefix\url{https://pos.sissa.it/398/376}

\bibitem{Maji2017a}
T.~Maji, D.~Chakrabarti,
  \href{http://link.aps.org/doi/10.1103/PhysRevD.95.074009}{{Transverse
  structure of a proton in a light-front quark-diquark model}}, Physical Review
  D 95~(7) (2017) 074009.
\newblock \href {http://dx.doi.org/10.1103/PhysRevD.95.074009}
  {\path{doi:10.1103/PhysRevD.95.074009}}.
\newline\urlprefix\url{http://link.aps.org/doi/10.1103/PhysRevD.95.074009}

\bibitem{Pasquini2008}
B.~Pasquini, S.~Cazzaniga, S.~Boffi,
  \href{https://link.aps.org/doi/10.1103/PhysRevD.78.034025}{{Transverse
  momentum dependent parton distributions in a light-cone quark model}},
  Physical Review D - Particles, Fields, Gravitation and Cosmology 78~(3)
  (2008) 034025.
\newblock \href {http://dx.doi.org/10.1103/PhysRevD.78.034025}
  {\path{doi:10.1103/PhysRevD.78.034025}}.
\newline\urlprefix\url{https://link.aps.org/doi/10.1103/PhysRevD.78.034025}

\bibitem{Musch2011}
B.~U. Musch, P.~H{\"{a}}gler, J.~W. Negele, A.~Sch{\"{a}}fer, {Exploring quark
  transverse momentum distributions with lattice QCD}, Physical Review D -
  Particles, Fields, Gravitation and Cosmology 83~(9) (2011) 1--38.
\newblock \href {http://arxiv.org/abs/1011.1213} {\path{arXiv:1011.1213}},
  \href {http://dx.doi.org/10.1103/PhysRevD.83.094507}
  {\path{doi:10.1103/PhysRevD.83.094507}}.

\bibitem{Musch2012}
B.~U. Musch, P.~H{\"{a}}gler, M.~Engelhardt, J.~W. Negele, A.~Sch{\"{a}}fer,
  \href{https://link.aps.org/doi/10.1103/PhysRevD.85.094510}{{Sivers and
  Boer-Mulders observables from lattice QCD}}, Physical Review D 85~(9) (2012)
  094510.
\newblock \href {http://dx.doi.org/10.1103/PhysRevD.85.094510}
  {\path{doi:10.1103/PhysRevD.85.094510}}.
\newline\urlprefix\url{https://link.aps.org/doi/10.1103/PhysRevD.85.094510}

\bibitem{Ji:2014hxa}
X.~Ji, P.~Sun, X.~Xiong, F.~Yuan, {Soft factor subtraction and transverse
  momentum dependent parton distributions on the lattice}, Phys. Rev. D 91
  (2015) 074009.
\newblock \href {http://arxiv.org/abs/1405.7640} {\path{arXiv:1405.7640}},
  \href {http://dx.doi.org/10.1103/PhysRevD.91.074009}
  {\path{doi:10.1103/PhysRevD.91.074009}}.

\bibitem{Yoon2017}
B.~Yoon, M.~Engelhardt, R.~Gupta, T.~Bhattacharya, J.~R. Green, B.~U. Musch,
  J.~W. Negele, A.~V. Pochinsky, A.~Sch{\"{a}}fer, S.~N. Syritsyn,
  \href{https://link.aps.org/doi/10.1103/PhysRevD.96.094508}{{Nucleon
  transverse momentum-dependent parton distributions in lattice QCD:
  Renormalization patterns and discretization effects}}, Physical Review D
  96~(9) (2017) 094508.
\newblock \href {http://dx.doi.org/10.1103/PhysRevD.96.094508}
  {\path{doi:10.1103/PhysRevD.96.094508}}.
\newline\urlprefix\url{https://link.aps.org/doi/10.1103/PhysRevD.96.094508}

\bibitem{Constantinou:2020hdm}
M.~Constantinou, et~al., {Parton distributions and lattice-QCD calculations:
  Toward 3D structure}, Prog. Part. Nucl. Phys. 121 (2021) 103908.
\newblock \href {http://arxiv.org/abs/2006.08636} {\path{arXiv:2006.08636}},
  \href {http://dx.doi.org/10.1016/j.ppnp.2021.103908}
  {\path{doi:10.1016/j.ppnp.2021.103908}}.

\bibitem{Shi:2018zqd}
C.~Shi, I.~C. Clo\"et, {Intrinsic Transverse Motion of the
  Pion\textquoteright{}s Valence Quarks}, Phys. Rev. Lett. 122~(8) (2019)
  082301.
\newblock \href {http://arxiv.org/abs/1806.04799} {\path{arXiv:1806.04799}},
  \href {http://dx.doi.org/10.1103/PhysRevLett.122.082301}
  {\path{doi:10.1103/PhysRevLett.122.082301}}.

\bibitem{Shi:2020pqe}
C.~Shi, K.~Bednar, I.~C. Clo\"et, A.~Freese, {Spatial and Momentum Imaging of
  the Pion and Kaon}, Phys. Rev. D 101~(7) (2020) 074014.
\newblock \href {http://arxiv.org/abs/2003.03037} {\path{arXiv:2003.03037}},
  \href {http://dx.doi.org/10.1103/PhysRevD.101.074014}
  {\path{doi:10.1103/PhysRevD.101.074014}}.

\bibitem{Vary2010}
J.~P. Vary, H.~Honkanen, J.~Li, P.~Maris, S.~J. Brodsky, A.~Harindranath, G.~F.
  {De Teramond}, P.~Sternberg, E.~G. Ng, C.~Yang,
  \href{https://link.aps.org/doi/10.1103/PhysRevC.81.035205}{{Hamiltonian
  light-front field theory in a basis function approach}}, Physical Review C -
  Nuclear Physics 81~(3) (2010) 035205.
\newblock \href {http://arxiv.org/abs/0905.1411} {\path{arXiv:0905.1411}},
  \href {http://dx.doi.org/10.1103/PhysRevC.81.035205}
  {\path{doi:10.1103/PhysRevC.81.035205}}.
\newline\urlprefix\url{https://link.aps.org/doi/10.1103/PhysRevC.81.035205}

\bibitem{Maris2013}
P.~Maris, P.~Wiecki, Y.~Li, X.~Zhao, J.~P. Vary, {Bound state calculations in
  QED and QCD using basis light-front quantization}, Acta Phys. Polon. Supp. 6
  (2013) 321--326.
\newblock \href {http://dx.doi.org/10.5506/APhysPolBSupp.6.321}
  {\path{doi:10.5506/APhysPolBSupp.6.321}}.

\bibitem{Zhao2014}
X.~Zhao, H.~Honkanen, P.~Maris, J.~P. Vary, S.~J. Brodsky,
  \href{https://linkinghub.elsevier.com/retrieve/pii/S0370269314005875}{{Electron
  g-2 in Light-front Quantization}}, Physics Letters, Section B: Nuclear,
  Elementary Particle and High-Energy Physics 737 (2014) 65--69.
\newblock \href {http://arxiv.org/abs/1402.4195} {\path{arXiv:1402.4195}},
  \href {http://dx.doi.org/10.1016/j.physletb.2014.08.020}
  {\path{doi:10.1016/j.physletb.2014.08.020}}.
\newline\urlprefix\url{https://linkinghub.elsevier.com/retrieve/pii/S0370269314005875}

\bibitem{Hu2021}
Z.~Hu, S.~Xu, C.~Mondal, X.~Zhao, J.~P. Vary,
  \href{https://doi.org/10.1103/PhysRevD.103.036005}{{Transverse structure of
  electron in momentum space in basis light-front quantization}}, Physical
  Review D 103~(3) (2021) 36005.
\newblock \href {http://dx.doi.org/10.1103/PhysRevD.103.036005}
  {\path{doi:10.1103/PhysRevD.103.036005}}.
\newline\urlprefix\url{https://doi.org/10.1103/PhysRevD.103.036005}

\bibitem{Xu2021}
S.~Xu, C.~Mondal, J.~Lan, X.~Zhao, Y.~Li, J.~P. Vary, K.~Fu, S.~Xu, Z.~Hu,
  X.~Zhao, J.~P. Vary, \href{http://arxiv.org/abs/2109.12921
  https://link.aps.org/doi/10.1103/PhysRevD.104.094036}{{Nucleon structure from
  basis light-front quantization}}, Physical Review D 104~(9) (2021) 094036.
\newblock \href {http://arxiv.org/abs/2109.12921} {\path{arXiv:2109.12921}},
  \href {http://dx.doi.org/10.1103/PhysRevD.104.094036}
  {\path{doi:10.1103/PhysRevD.104.094036}}.
\newline\urlprefix\url{http://arxiv.org/abs/2109.12921
  https://link.aps.org/doi/10.1103/PhysRevD.104.094036}

\bibitem{Mondal2020}
C.~Mondal, S.~Xu, J.~Lan, X.~Zhao, Y.~Li, D.~Chakrabarti, J.~P. Vary,
  \href{https://link.aps.org/doi/10.1103/PhysRevD.102.016008}{{Proton structure
  from a light-front Hamiltonian}}, Physical Review D 102~(1) (2020) 016008.
\newblock \href {http://arxiv.org/abs/1911.10913} {\path{arXiv:1911.10913}},
  \href {http://dx.doi.org/10.1103/PhysRevD.102.016008}
  {\path{doi:10.1103/PhysRevD.102.016008}}.
\newline\urlprefix\url{https://link.aps.org/doi/10.1103/PhysRevD.102.016008}

\bibitem{Liu2022}
Y.~Liu, S.~Xu, C.~Mondal, X.~Zhao, J.~P. Vary,
  \href{http://arxiv.org/abs/2202.00985}{{Angular momentum and generalized
  parton distributions for the proton with basis light-front quantization}}
  (2022) 1--14\href {http://arxiv.org/abs/2202.00985}
  {\path{arXiv:2202.00985}}.
\newline\urlprefix\url{http://arxiv.org/abs/2202.00985}

\bibitem{Lepage1980}
G.~P. Lepage, S.~J. Brodsky, {Exclusive processes in perturbative quantum
  chromodynamics}, Physical Review D 22~(9) (1980) 2157--2198.
\newblock \href {http://dx.doi.org/10.1103/PhysRevD.22.2157}
  {\path{doi:10.1103/PhysRevD.22.2157}}.

\bibitem{Pauli1985}
H.~C. Pauli, S.~J. Brodsky, {Solving field theory in one space and one time
  dimension}, Physical Review D 32~(8) (1985) 1993--2000.
\newblock \href {http://dx.doi.org/10.1103/PhysRevD.32.1993}
  {\path{doi:10.1103/PhysRevD.32.1993}}.

\bibitem{Brodsky2001a}
S.~J. Brodsky, M.~Diehl, D.~S. Hwang, \href{http://arxiv.org/abs/hep-ph/0009254
  http://dx.doi.org/10.1016/S0550-3213(00)00695-7
  https://linkinghub.elsevier.com/retrieve/pii/S0550321300006957}{{Light-cone
  wavefunction representation of deeply virtual Compton scattering}}, Nuclear
  Physics B 596~(1-2) (2001) 99--124.
\newblock \href {http://arxiv.org/abs/0009254} {\path{arXiv:0009254}}, \href
  {http://dx.doi.org/10.1016/S0550-3213(00)00695-7}
  {\path{doi:10.1016/S0550-3213(00)00695-7}}.
\newline\urlprefix\url{http://arxiv.org/abs/hep-ph/0009254
  http://dx.doi.org/10.1016/S0550-3213(00)00695-7
  https://linkinghub.elsevier.com/retrieve/pii/S0550321300006957}

\bibitem{Soper1972}
D.~E. Soper,
  \href{https://link.aps.org/doi/10.1103/PhysRevD.5.1956}{{Infinite-momentum
  helicity states}}, Physical Review D 5~(8) (1972) 1956--1962.
\newblock \href {http://dx.doi.org/10.1103/PhysRevD.5.1956}
  {\path{doi:10.1103/PhysRevD.5.1956}}.
\newline\urlprefix\url{https://link.aps.org/doi/10.1103/PhysRevD.5.1956}

\bibitem{Brodsky2015}
S.~J. Brodsky, G.~F. de~T{\'{e}}ramond, H.~G. Dosch, J.~Erlich, {Light-front
  holographic QCD and emerging confinement}, Physics Reports 584 (2015) 1--105.
\newblock \href {http://arxiv.org/abs/1407.8131} {\path{arXiv:1407.8131}},
  \href {http://dx.doi.org/10.1016/j.physrep.2015.05.001}
  {\path{doi:10.1016/j.physrep.2015.05.001}}.

\bibitem{Li2017}
Y.~Li, P.~Maris, J.~P. Vary,
  \href{http://link.aps.org/doi/10.1103/PhysRevD.96.016022}{{Quarkonium as a
  relativistic bound state on the light front}}, Physical Review D 96~(1)
  (2017) 016022.
\newblock \href {http://arxiv.org/abs/1704.06968} {\path{arXiv:1704.06968}},
  \href {http://dx.doi.org/10.1103/PhysRevD.96.016022}
  {\path{doi:10.1103/PhysRevD.96.016022}}.
\newline\urlprefix\url{http://link.aps.org/doi/10.1103/PhysRevD.96.016022}

\bibitem{Li2016}
Y.~Li, P.~Maris, X.~Zhao, J.~P. Vary,
  \href{https://linkinghub.elsevier.com/retrieve/pii/S0370269316301472}{{Heavy
  quarkonium in a holographic basis}}, Physics Letters, Section B: Nuclear,
  Elementary Particle and High-Energy Physics 758~(Yang Li) (2016) 118--124.
\newblock \href {http://arxiv.org/abs/1509.07212} {\path{arXiv:1509.07212}},
  \href {http://dx.doi.org/10.1016/j.physletb.2016.04.065}
  {\path{doi:10.1016/j.physletb.2016.04.065}}.
\newline\urlprefix\url{https://linkinghub.elsevier.com/retrieve/pii/S0370269316301472}

\bibitem{Wiecki2015}
P.~Wiecki, Y.~Li, X.~Zhao, P.~Maris, J.~P. Vary,
  \href{https://link.aps.org/doi/10.1103/PhysRevD.91.105009}{{Basis light-front
  quantization approach to positronium}}, Physical Review D - Particles,
  Fields, Gravitation and Cosmology 91~(10) (2015) 105009.
\newblock \href {http://arxiv.org/abs/1404.6234} {\path{arXiv:1404.6234}},
  \href {http://dx.doi.org/10.1103/PhysRevD.91.105009}
  {\path{doi:10.1103/PhysRevD.91.105009}}.
\newline\urlprefix\url{https://link.aps.org/doi/10.1103/PhysRevD.91.105009}

\bibitem{Goeke2005}
K.~Goeke, A.~Metz, M.~Schlegel, {Parameterization of the quark-quark correlator
  of a spin-1/2 hadron}, Physics Letters, Section B: Nuclear, Elementary
  Particle and High-Energy Physics 618~(1-4) (2005) 90--96.
\newblock \href {http://arxiv.org/abs/0504130} {\path{arXiv:0504130}}, \href
  {http://dx.doi.org/10.1016/j.physletb.2005.05.037}
  {\path{doi:10.1016/j.physletb.2005.05.037}}.

\bibitem{Meissner2007}
S.~Mei{\ss}ner, A.~Metz, K.~Goeke, \href{http://arxiv.org/abs/hep-ph/0703176
  http://dx.doi.org/10.1103/PhysRevD.76.034002}{{Relations between generalized
  and transverse momentum dependent parton distributions}}, Physical Review D -
  Particles, Fields, Gravitation and Cosmology 76~(3) (2007) 1--25.
\newblock \href {http://arxiv.org/abs/0703176} {\path{arXiv:0703176}}, \href
  {http://dx.doi.org/10.1103/PhysRevD.76.034002}
  {\path{doi:10.1103/PhysRevD.76.034002}}.
\newline\urlprefix\url{http://arxiv.org/abs/hep-ph/0703176
  http://dx.doi.org/10.1103/PhysRevD.76.034002}

\bibitem{Meissner2009}
S.~Mei{\ss}ner, A.~Metz, M.~Schlegel, {Generalized parton correlation functions
  for a spin-1/2 hadron}, Journal of High Energy Physics 2009~(8) (2009) 0--39.
\newblock \href {http://arxiv.org/abs/0906.5323} {\path{arXiv:0906.5323}},
  \href {http://dx.doi.org/10.1088/1126-6708/2009/08/056}
  {\path{doi:10.1088/1126-6708/2009/08/056}}.

\bibitem{Lorce2011b}
C.~Lorc{\'{e}}, B.~Pasquini, {Origin of model relations among
  transverse-momentum dependent parton distributions}, Physical Review D -
  Particles, Fields, Gravitation and Cosmology 84~(3) (2011) 1--18.
\newblock \href {http://arxiv.org/abs/1104.5651} {\path{arXiv:1104.5651}},
  \href {http://dx.doi.org/10.1103/PhysRevD.84.034039}
  {\path{doi:10.1103/PhysRevD.84.034039}}.

\bibitem{Goeke2003}
K.~Goeke, A.~Metz, P.~V. Pobylitsa, M.~V. Polyakov, {Lorentz invariance
  relations among parton distributions revisited}, Physics Letters, Section B:
  Nuclear, Elementary Particle and High-Energy Physics 567~(1-2) (2003) 27--30.
\newblock \href {http://dx.doi.org/10.1016/S0370-2693(03)00870-0}
  {\path{doi:10.1016/S0370-2693(03)00870-0}}.

\bibitem{Mulders1995}
P.~J. Mulders, R.~D. Tangerman, {The Complete tree level result up to order 1/Q
  for polarized deep inelastic leptoproduction}, Nucl. Phys. B 461 (1996)
  197--237.
\newblock \href {http://arxiv.org/abs/9510301} {\path{arXiv:9510301}}, \href
  {http://dx.doi.org/10.1016/0550-3213(95)00632-X}
  {\path{doi:10.1016/0550-3213(95)00632-X}}.

\bibitem{Bacchetta2000}
A.~Bacchetta, M.~Boglione, A.~Henneman, P.~J. Mulders, {Bounds on transverse
  momentum dependent distribution and fragmentation functions}, Physical Review
  Letters 85~(4) (2000) 712--715.
\newblock \href {http://dx.doi.org/10.1103/PhysRevLett.85.712}
  {\path{doi:10.1103/PhysRevLett.85.712}}.

\bibitem{Scimemi2020}
I.~Scimemi, A.~Vladimirov, {Non-perturbative structure of semi-inclusive
  deep-inelastic and Drell-Yan scattering at small transverse momentum},
  Journal of High Energy Physics 2020~(6).
\newblock \href {http://dx.doi.org/10.1007/JHEP06(2020)137}
  {\path{doi:10.1007/JHEP06(2020)137}}.

\bibitem{Anselmino2014}
M.~Anselmino, M.~Boglione, J.~O. {Gonzalez H.}, S.~Melis, A.~Prokudin,
  {Unpolarised transverse momentum dependent distribution and fragmentation
  functions from SIDIS multiplicities}, Journal of High Energy Physics
  2014~(4).
\newblock \href {http://dx.doi.org/10.1007/JHEP04(2014)005}
  {\path{doi:10.1007/JHEP04(2014)005}}.

\bibitem{Bacchetta2017}
A.~Bacchetta, F.~Delcarro, C.~Pisano, M.~Radici, A.~Signori,
  \href{http://arxiv.org/abs/1703.10157
  http://dx.doi.org/10.1007/JHEP06(2017)081
  http://link.springer.com/10.1007/JHEP06(2017)081}{{Extraction of partonic
  transverse momentum distributions from semi-inclusive deep-inelastic
  scattering, Drell-Yan and Z-boson production}}, Journal of High Energy
  Physics 2017~(6) (2017) 81.
\newblock \href {http://arxiv.org/abs/1703.10157} {\path{arXiv:1703.10157}},
  \href {http://dx.doi.org/10.1007/JHEP06(2017)081}
  {\path{doi:10.1007/JHEP06(2017)081}}.
\newline\urlprefix\url{http://arxiv.org/abs/1703.10157
  http://dx.doi.org/10.1007/JHEP06(2017)081
  http://link.springer.com/10.1007/JHEP06(2017)081}

\bibitem{Bertone2019}
V.~Bertone, I.~Scimemi, A.~Vladimirov, {Extraction of unpolarized quark
  transverse momentum dependent parton distributions from Drell-Yan/Z-boson
  production}, Journal of High Energy Physics 2019~(6).
\newblock \href {http://arxiv.org/abs/1902.08474} {\path{arXiv:1902.08474}},
  \href {http://dx.doi.org/10.1007/JHEP06(2019)028}
  {\path{doi:10.1007/JHEP06(2019)028}}.

\bibitem{Signori2013}
A.~Signori, A.~Bacchetta, M.~Radici, G.~Schnell, {Investigations into the
  flavor dependence of partonic transverse momentum}, Journal of High Energy
  Physics 2013~(11).
\newblock \href {http://arxiv.org/abs/1309.3507} {\path{arXiv:1309.3507}},
  \href {http://dx.doi.org/10.1007/JHEP11(2013)194}
  {\path{doi:10.1007/JHEP11(2013)194}}.

\bibitem{Bacchetta2008a}
A.~Bacchetta, D.~Boer, M.~Diehl, P.~J. Mulders,
  \href{http://stacks.iop.org/1126-6708/2008/i=08/a=023?key=crossref.8a524230f6fa4a679073c5c50c534976}{{Matches
  and mismatches in the descriptions of semi-inclusive processes at low and
  high transverse momentum}}, Journal of High Energy Physics 2008~(08) (2008)
  023--023.
\newblock \href {http://dx.doi.org/10.1088/1126-6708/2008/08/023}
  {\path{doi:10.1088/1126-6708/2008/08/023}}.
\newline\urlprefix\url{http://stacks.iop.org/1126-6708/2008/i=08/a=023?key=crossref.8a524230f6fa4a679073c5c50c534976}

\end{thebibliography}
\end{document}